\documentclass[published]{JHEP3} 
\JHEP{10(2003)035}
\received{September 9, 2003}
\accepted{October 15, 2003}

\skip\footins = 1\bigskipamount plus 2pt minus 4pt

\usepackage{multicol}
\usepackage{supertabular,theorem,marvosym}

\newcommand{\diag}{\rm diag}
\newcommand{\leb}{\left |}
\newcommand{\rib}{\right |}

\def\mrm{\mathrm}
\def\be{\begin{equation}}
\def\ee{\end{equation}}
\def\bea{\begin{eqnarray}}
\def\eea{\end{eqnarray}}

\newcommand\U{\mathop{\rm {}U}\nolimits} 
\newcommand\SU{\mathop{\rm SU}\nolimits} 
\newcommand\SO{\mathop{\rm SO}\nolimits} 

\title{Hierarchical neutrino mass matrices, CP violation and leptogenesis}

\author{Liliana Velasco-Sevilla\\
Theoretical Physics, University of Oxford\\
1 Keble Road, Oxford OX1 3NP, U.K.\\   
E-mail: \email{velasco@thphys.ox.ac.uk}}

\abstract{In this work we study examples of hierarchical neutrino mass
  matrices inspired by family symmetries, compatible with experiments
  on neutrino oscillations, and for which there is a connection among
  the low energy CP violation phase associated to neutrino
  oscillations, the phases appearing in the amplitude of neutrinoless
  double beta decay, and the phases relevant for leptogenesis. In
  particular, we determine the predictions from a texture based on an
  underlying ${\rm SU}(3)$ family symmetry together with a GUT
  symmetry, and a strong hierarchy for the masses of the heavy right
  handed Majorana masses. We also give some examples of inverted
  hierarchies of neutrino masses, which may be motivated in the
  context of ${\rm U}(1)$ family symmetries.}

\begin{document}

\section{Introduction}\label{section1}

This work is motivated by the study, in the context of models
accommodating the masses of leptons, of the correlation among the
parameters associated to CP violation, appearing in neutrino
oscillations, CP violation in leptogenesis and CP violation appearing
in neutrinoless double beta decay ($\beta\beta_{\nu_0}$). In
particular we study the kind of models presented in~\cite{Ross:2002fb}
(which are based on an underlying $\SU(3)$ family symmetry together
with a GUT symmetry and a strong hierarchy for the \pagebreak[3]
masses of the heavy right handed Majorana masses) and also models
giving inverted hierarchies for neutrino masses, which can be
understood in the context of a $\U(1)$ family symmetry.

Recently there have appeared some
studies~\cite{King:2002qh}--\cite{Davidson:2002qv} of a possible
connection between low energy phases and those phases relevant for
leptogenesis, motivated by the fact that leptogenesis is a very
attractive candidate in explaining the baryon asymmetry of the
universe (BAU) and also by the present information coming from
neutrino oscillation experiments. In the leptogenesis scenario a $B-L$
asymmetry is produced from the decay of the heavy right-handed
Majorana neutrinos, $N_j$. This asymmetry is parameterized in terms of
asymmetry parameters, $\varepsilon_j$, which can be expressed in terms
of the heavy right handed Majorana masses and the Yukawa couplings for
neutrinos. Therefore, in the context of models describing the correct
neutrino mass splittings and mixings, it is natural to look for such a
connection and for a correlation between the sign of the baryon number
of the universe and the strength of the CP violation in neutrino
oscillations. Such correlations are not a general feature of the
models explaining neutrino oscillations and in some cases are quite
model dependent. Nevertheless, if there are plausible models for
family symmetries explaining not only neutrino oscillations but also
the low energy parameters of quarks, it is interesting to examine
whether or not these models are compatible with the leptogenesis
scenario and whether or not it is possible to establish correlations
among CP asymmetries appearing in leptogenesis, the parameters of CP
violation in neutrino oscillations and the phases appearing in the
neutrinoless double beta decay. The relevance of neutrinoless double
decay processes is that not only the Majorana nature of neutrinos can
be unveiled, since their amplitude is proportional to an average mass
containing the Majorana phases of neutrinos, but also that the scale
of neutrino masses may be determined.

Neutrino oscillation experiments have provided evidence for
non-vanishing neutrino masses and mixings. Two years ago there were
yet considered many possibilities to explain the solar neutrino
mixing, now the recent results from KamLAND~\cite{Eguchi:2002dm}
indicate that the most favoured solution is the MSW LMA solution. This
information and that coming from the atmospheric mixing
analyses~\cite{pdholsmsno02}--\cite{bamawhisno02},
which may be summarized as follows
\bea
\label{expcons}
\tan^2{\theta _{\rm atm}}&=&1^{-0.57}_{+1.3}\,,
\qquad 
\tan^2{\theta _{\rm sol}}=(4^{-1.7}_{+3.9})\times 10^{-1}\,,
\nonumber\\ 
\tan^2{\theta _{\rm rct}}&<&5.5 \times 10^{-2}\,,
\nonumber\\ 
\Delta m^2_{\rm sol}&\in &(5.1,9.7)\times 10^{-5}\,{\rm eV}^2\,, 
\qquad
\Delta m^2_{\rm atm}\in(1.2,4.8)\times 10^{-3}\,{\rm eV}^2\,,
\eea
has given us a definite point of departure from the Standard Model
(SM)\footnote{Here $\theta_{\mrm{atm}}$, $\theta_{\mrm{sol}}$,
    $\theta_{\mrm{rct}}$ are the mixing angles describing atmospheric,
    solar, and reactor neutrino oscillation experiments,
    respectively.} and thus a way to probe possible symmetries for
  leptons and quarks in an unified scheme.  With this information it
  has been possible to carry out {\it bottom-up} approach
  analyses~\cite{Lavignac:2002gf,Frigerio:2002rd,Frigerio:2002fb} in
  order to reconstruct the possible forms of the effective neutrino
  mass matrix. The most plausible forms consistent with data are: (i)
  \emph{hierarchical (canonical)}, which can be such that $m_1 \ll m_2
  \ll m_3$ or $m_1 \lesssim m_2 \ll m_3$, (ii) \emph{inverted
    hierarchical} $m_2 \gtrsim m_1 \gg m_3$ or (iii) \emph{degenerate}
  $m_1^2 \approx m_2^2 \approx m_3^2$.

\pagebreak[3] 

A direct way to re-construct the possible forms of the neutrino mass
matrix is by working in the \emph{flavour basis} in which the charged
lepton mass matrix is diagonal and the mixings appearing in the lepton
mixing matrix, $U_{\rm MNS}$ matrix are only due to
neutrinos. However, in trying to identify the possible broken
symmetries underlying the neutrino `puzzle', it is convenient to
extract the neutrino mass matrix in the \emph{symmetry basis}, in
which the patterns of the possible (broken) symmetries underlying the
leptons is reflected on the structure of the mass matrices. In this
basis, charged lepton and neutrino mass matrices may not be diagonal
and hence we can study the possible contributions of charged leptons
and neutrinos to the $U_{{\rm MNS}}$ matrix, given by different family
symmetries.

The work is organized as follows. In section~\ref{section2} we review
the basic structure of a family symmetry motivated by an
$\SU(3)_F\times \SO(10)_{\rm GUT}$ symmetry~\cite{Ross:2002fb} and the
general structure of $\U(1)_F$ family symmetries. In
section~\ref{section3}, we comment upon the diagonalization of
hierarchical mass matrices, which can be used for canonical and
inverted hierarchies, discussing the details in appendix~\ref{appA}. In this
section we also construct the $U_{\rm MNS}$ in terms of those matrices
diagonalizing neutrino and charged lepton mass matrices. In
section~\ref{section4} we derive the predictions for the CP violation
phase appearing in neutrino oscillations, $\delta_{\rm O}$, for two
different kinds of hierarchies of the neutrino mass matrix (one
producing the hierarchy $m_{\nu_3}\gg m_{\nu_2}\gg m_{\nu_1}$ and the
other producing the hierarchy $m_{\nu_2}\gtrsim m_{\nu_1}\gg
m_{\nu_3}$). We give an explicit realization of each one, motivated by
the $\SU(3)_F\times \SO(10)_{\rm GUT}$ and the $\U(1)_F$ family
symmetries, respectively. In section~\ref{section5} we determine that
these hierarchies are compatible with the leptogenesis scenario,
giving an approximate value of the baryon asymmetry produced, and
comment upon the connections between the phases appearing in neutrino
oscillations and the phases relevant for leptogenesis. In
section~\ref{section6} we determine the Majorana phases for the
hierarchies presented and comment further upon the relations to the
leptogenesis phase. We conclude with a summary and outlook.

\section{Family symmetries and symmetry basis}\label{section2}

The possibility of explaining the masses of quarks and leptons through
a set of symmetries containing the SM model has been widely
explored. Such explanation may be achieved within the context of more
fundamental theories such as String Theory, Grand Unified theories and
Flavour symmetries (those symmetries distinguishing between families)
also called \emph{horizontal symmetries}. We consider models in which
there is an underlying family symmetry and the neutrino masses are
given by the see-saw
mechanism~\cite{GellRamSla79}--\cite{Mohapatra:1980ia}.  When these
symmetries are broken they leave an imprint in the form of the mass
matrices appearing in the effective mass lagrangian. In the leptonic
sector, this has the form
\bea
\label{effectmasslang}
-L^{\ell}_m&=&\bar{\nu^o}_{L} m^{\nu}_D \nu^o_{R} 
+\frac{1}{2}\bar{\nu^c}_{R} M_R \nu_{R}+ \bar{l^o}_L m^ll^o_R +\mbox{h.c.} 
\label{effectmasslag}\\
&=&\frac{1}{2}\bar{n^c}_L m^{\nu}_{LL}n_L+ \bar{l^o}_{L}m^ll^o_{R}+h.c.\,,
\label{effmasslag2}
\eea
where $\nu^o_R$ labels the right-handed (R-H) Majorana neutrinos,
$l^o$ the charged leptons and we have assumed that the possible
Majorana mass term associated with the left-handed neutrinos,
$\nu^o_L$, vanishes. The state $n_L=(\nu^o_L,(\nu^o)^c_R)$, has an
effective mass $m^{\nu}_{LL}$ given, approximately, by the see-saw
formula~\cite{GellRamSla79}--\cite{Mohapatra:1980ia},
\be
\label{seesawfor}
m^{\nu\dagger}_{LL}\approx -m^{\nu}_D M^{-1}_R (m^{\nu}_D)^t\,.
\ee
Each matrix in eq.~(\ref{effectmasslag}) is not necessarily diagonal,
showing the patterns left by the broken family symmetries. This basis
defines what we call the \emph{symmetry basis}.

The most plausible
patterns~\cite{Lavignac:2002gf,Frigerio:2002rd,Frigerio:2002fb}
describing the effective neutrino mass matrix $m^{\nu}_{LL}$ are the
given by the following patterns, namely the hierarchical pattern
\be
{\mrm H}:\quad
m_{LL}^{\nu}=\pmatrix{
\epsilon^{\prime} &\epsilon &\epsilon\cr
\epsilon & 1&1\cr
\epsilon & 1&1}
\frac{\tilde{m}_{\nu}}{2}\,,
\ee
which produces the ordering $m_{\nu_3}\gg m_{\nu_2} > m_{\nu_1} $, the
inverted hierarchical patterns:
\bea
\label{plausinvhi}
{\mrm IH1}:\quad
m_{LL}^{\nu}&=&\pmatrix{
\epsilon^{\prime} &\frac{1}{\sqrt{2}} &\frac{1}{\sqrt{2}}\cr
\frac{1}{\sqrt{2}}&\epsilon &\epsilon \cr
\frac{1}{\sqrt{2}}&\epsilon &\epsilon }
\tilde{m}_{\nu}\,,
\qquad
{\mrm{IH2}}:\quad m_{LL}^{\nu}=\pmatrix{
1 &\epsilon &\epsilon\cr
\epsilon & \frac{1}{2}&\frac{1}{2}\cr
\epsilon & \frac{1}{2}&\frac{1}{2}}
\tilde{m}_{\nu}\,,
\nonumber\\ 
\tilde{m}_{\nu}&=&{\mathcal{O}}({\rm max}\{m_{\nu_3}, 
m_{\nu_2}, m_{\nu_1}\})\,,
\eea
which gives rise to the ordering $m_{\nu_3}\ll m_{\nu_1} \lesssim
m_{\nu_2} $, and, finally, the anarchical patterns, in which none of
the elements are related and may give rise to different spectra. Here
$\epsilon, \epsilon^{\prime}\ll 1$ and we have indicated only the
order of magnitude without specifying the order 1 coefficients.  The
hierarchies $H$ and $IH2$ can be realized in the flavour basis, where
the charged leptons are diagonal, reproducing the observed lepton
mixings and neutrino mass splittings. The inverted hierarchy $IH2$
typically gives a negligible mixing angle,
$\theta^{\nu}_{13}$~\cite{He:2002rv}. The hierarchy $IH1$ produces a
rather small angle $\theta^{\nu}_{12}$, unless there is a fine tunning
in the parameters. The three hierarchies may be realized in suitable
models accommodating neutrinos and leptons, where the mixing angles
arising from the diagonalization of $m^{\nu}_{LL}$ receive
contributions coming from the diagonalization of $m^l$ and especially
those from $\theta_{\rm rct}$, as we shall shortly see in
section~\ref{predcanhier}.

Although there have been several proposals of symmetries capable of
accommodating these different
possibilities~\cite{Caldwell:1995vi,King:2000ce,Babu:2001cv,He:2002rv,Ohlsson:2002na,Altarelli:2002sg},
only some of them~\cite{King:2000ce,Ohlsson:2002na,Ross:2002fb} have
been given in the context of a more general model for quarks and
leptons. We are interested in knowing whether in models of this kind,
there are correlations among the phases appearing in neutrino
oscillations, neutrinoless double beta decay and leptogenesis. To be
more concrete, we would like to analyze predictions for: 
\begin{itemize}
\item[(I)] A class of hierarchies of the type $H1$ and detailed
  predictions for an specific example that may be realized in the
  context of an $\SU(3)_F\times \SO(10)_{GUT}$ symmetry,
\item[(II)] a hierarchy of the type $IH2$ in the context of $\U(1)_F$
  symmetries.
\end{itemize}

Before we present the predictions for the CP violation phases, and in
order to motivate the hierarchies $H$, $IH1$ and $IH2$ in the context
of family symmetries, we review the general features of the family
symmetries $\SU(3)_F$ and $U_F$ that we consider.

\paragraph{The case of a texture inspired in a $\SU(3)_F\times \SO(10)$ symmetry.} In~\cite{Ross:2002fb} we have considered a texture for quarks and
leptons inspired by a supersymmetric $\SU(3)_F\times \SO(10)_{\rm
  GUT}$\footnote{The group $\SU(3)$ as a family symmetry has been
  broadly considered in the
  past~\cite{Chkareuli:1980gy}--\cite{Chkareuli:2001dq}.}  family
symmetry where we have parameterized the Yukawa matrices for fermions
by
\be
\label{massmatpar}
Y^f=\frac{m^f_D}{(m^f_D)_{(3,3)}}=\pmatrix{
\varepsilon ^{8} & \varepsilon ^{3}(z+(x+y)\varepsilon) & \varepsilon ^{3}(z
+(x-y)\varepsilon) \cr
-\varepsilon ^{3}(z+(x+y)\varepsilon) & \varepsilon ^{2}(a_{f}w+u\varepsilon) 
& \varepsilon ^{2}(a_{f}w-u\varepsilon ) \cr
-\varepsilon ^{3}(z +(x-y)\varepsilon) & \varepsilon ^{2}
(a_{f}w-u\varepsilon) & 1},
\ee
here the index $f$ denotes the kind of fermions: $f=\nu, u, d, l$. In
the context of a supersymmetric theory of fermion masses we may
construct the Yukawa superpotential based on the terms allowed by the
symmetry; for example for eq.~(\ref{massmatpar}), the dominant terms
are given by the operators~\cite{kingross}
\be
\label{termssu3}
\left(\frac{1}{M_3^2}\psi_i \phi^i_{3}\psi^c_j\phi^j_{3}+ 
\frac{1}{M^2}\psi_i\phi^i_{23}\psi^c_j\phi^j_{23}\right)H_{\alpha}\,,
\quad i,j=1,2,3
\ee
where $\psi_i$ and $\psi^c_i$ are left-handed quarks and leptons
($\psi_i\in(Q_i, L_i)$, $\psi^c_i\in(U^c_i,D^c_i,E^c_i, N^c_i)$) which
transform, respectively, as the fundamental and its conjugate
representation of $\SU(3)$. $\phi^i_3$ and $\phi^i_{23}$ are scalar
anti-triplet fields responsible for the symmetry breaking of $\SU(3)$
with $H_{\alpha}$ being the two Higgs doublets of the Minimal
Supersymmetric Standard Model (MSSM), which are singlets under the
$\SU(3)$ family symmetry. When the symmetry is spontaneously broken
the scalar field $\phi_3$ acquires a vev (vacuum expectation value)
such that the term $\frac{1}{M_3^2}\psi_i
\phi^i_{3}\psi^c_j\phi^j_{3}$ produces a Yukawa coupling only for the
$(3,3)$ entry. The vev of $\phi_{23}$ produces entries $(2,2)$,
$(2,3)$ and $(3,2)$ at order $\varepsilon^2$, through the second
operator in eq.~(\ref{termssu3}), where $\varepsilon={\langle
  \phi_{23} \rangle}/{M}$. The other entries of eq.~(\ref{massmatpar})
are generated through higher dimensional operators.

The coefficients $z$, $x$, $y$, $w$ and $u$ are complex numbers of
order 1 and can be fitted to reproduce the values of the atmospheric,
solar and reactor neutrino mixing angles~\cite{Ross:2002fb}. The
coefficient $a_f$ is a coefficient that depends on the nature of
fermion coupling to the Higgs boson. This is described as an effective
120 or 126 $\SO(10)$ representation, coming from a coupling to
$H_{10}$ and $\Sigma _{45}$.  Here $\Sigma$ acquires a vev given by
$\langle\Sigma \rangle=B-L+\kappa T_{R,3}$, where $B$ and $L$ are the baryon
number and lepton number operators, respectively, and $T_{R,3}$ is the
third component of isospin in $\SO(10)$. We have two solutions
\begin{eqnarray} 
\kappa &=& 0\,,\qquad a_{l}=-3\,,\quad a_{\nu }=-3\,,
\nonumber\\ 
\kappa &=& 2\,,\qquad a_{l}=+3\,,\quad a_{\nu }=0
\end{eqnarray}
but only with the last one we obtain large atmospheric and solar
mixings~\cite{Ross:2002fb}.

\paragraph{Inverted hierarchies in the context of a $\U(1)_F$ symmetries}

$\U(1)_F$ family symmetries provide a convenient way of organizing the
hierarchies within the Yukawa matrices for fermions (see for
example~\cite{Ibanez:1994ig, King:1999mb}). We consider here a
$\U(1)_F$, which is broken by a set of singlets $\theta$ and
$\bar{\theta}$, such that the breaking scale, $M_Y$, is set by
$\theta=\bar{\theta}$, where the vevs are acquired along a `D-flat
direction'.  The general idea is that at tree-level the $\U(1)_F$
symmetry only allows one Yukawa coupling, generally for the third
family, due to the dominance of the top quark Yukawa coupling.
Smaller Yukawa couplings may be generated effectively from higher
dimension non-renormalizable operators.  Such operators correspond to
insertions of $\theta$ and $\bar{\theta}$, and hence to powers of the
expansion parameter $\epsilon={\langle \theta \rangle}/{M_Y}$. The
power of the expansion parameter is controlled by the $\U(1)_F$
charges of the particular operator. The relevant fields for leptons
are the lepton doublets $L_i$, the charge conjugated right-handed
neutrinos and charged leptons $(\nu^o_R)^c$, $(l^o)^c$, the up-type
Higgs doublet $H_u$ and a single scalar field, $H_M$.  The vev of the
latter is responsible for giving mass to the heavy Majorana fields. We
may denote the charges of these fields by, $l_i$, $n_i$, $e_i$, $h_u$
and $h_M$ respectively, and, within this convention, the Yukawa
couplings for neutrinos and charged lepton may be written in the form
\be
\label{yukawacoupne}
(Y^{\nu})_{ij}=\epsilon^{|l_i+n_j|}\,,
\qquad 
(Y^e)_{ij}=\epsilon^{|l_i+e_j|}\,,
\ee 
where we have re-absorbed the charge $h_u$ into the definition of the
lepton charges $l_i$. The heavy right handed neutrino mass matrix is
given by
\be
\label{rightmassu1}
(M_R)_{ij}=\epsilon^{|n_i+n_j+h_M|}\langle H_M\rangle\,.
\ee 
The assignment of charges under the $\U(1)_F$ is constrained to
reproduce lepton masses and mixings. The mass matrices for quarks can
also be expressed in terms of $\U(1)_F$ symmetries, whose Yukawa
couplings would be of the form
$(Y^q)_{i,j}=\epsilon^{|(q_L)_i+(q_R)_j|}$, where $q_i$ are the
$\U(1)_F$ charges. However, as we have seen in~\cite{rrrv}, the kind
of predictions for elements of the CKM matrix would be disfavoured,
according to precision tests against the experimental measurements
contributing to the CKM matrix. Nevertheless, if the $\U(1)_F$
symmetries are realized in the context of a GUT theory, it is possible
to improve their predictions (see for
example~\cite{Altarelli:1999dg}).

\section{$U_{\rm MNS}$ in the symmetry basis}\label{section3}

The mixing matrices for quarks and leptons are described in terms of a
unitary $3\times 3$ matrix which in general can be parameterized in
terms of $3$ angles and $6$ phases, namely,
\be
\label{convmajph}
{\rm diag}(e^{i\sigma_3},e^{i\sigma_4},e^{i\sigma_5}) U
{\rm diag}(e^{i\sigma_1},1,e^{i\sigma_2})\equiv P^{\prime}(\sigma)U P(\sigma)
\ee
where $U$ contains a single phase, $\delta$.  For quarks the five
$\sigma_m$ phases may be absorbed into the re-definition of quark
fields but for leptons, due to the Majorana nature of neutrinos, we
may absorb just three phases into the re-definition of the charged
lepton fields.  Thus, we are left with \emph{two of the} $\sigma_m$
phases which may, in turn, be associated with the effective Majorana
neutrino mass matrix, $ (m^{\nu}_{LL})_{\rm diag}={\rm diag}(m_{\nu_1}
e^{-2i\sigma_1},m_{\nu_2} ,m_{\nu_3}e^{-2i\sigma_2}) $, or may remain
in the mixing matrix:
\be
\label{umnsandpsigma}
U_{\rm MNS}=UP(\sigma)\,.
\ee
The latter convention, adopted here, for $U_{MNS}$ requires three
mixing angles and three CP violation phases. One of these phases is
analogous to the case of quarks, $\delta$, which is often called the
\emph{Dirac CP violation} phase, and the other two, $\sigma_1$ and
$\sigma_2$, are the \emph{ Majorana CP violation} phases.  For the
case of leptons and quarks, the standard parameterization for $U$
adopted here is:
\bea
\label{stparmixm}
U&=&R_{23}P(-\delta,1,1)R_{13}P(\delta,1,1)R_{12}
\nonumber\\
&=&\pmatrix{
c_{13}c_{12}&s_{12}c_{13}&s_{13}e^{-i\delta}\cr
-s_{12}c_{23}-s_{23}s_{13}c_{12}e^{i\delta}&
c_{23}c_{12}-s_{23}s_{13}s_{12}e^{i\delta}&s_{23}c_{13}\cr
s_{23}s_{12}-s_{13}c_{23}c_{12}e^{i\delta}&-s_{23}c_{12}
-s_{13}s_{12}c_{23}e^{i\delta}&c_{23}c_{13}}
\eea
where $c_{ij}\equiv \cos\theta_{ij}$, $s_{ij}\equiv \sin\theta_{ij}$.
Here the mixing angles vary between 0 and $\pi/2$ and $\delta$ varies
between 0 and $2\pi$. Thus $U$ may be expressed as the product of the
matrices $R_{ij}$, rotations in the $ij$ plane, such that
$(R_{ij})_{ij}=s_{ij}$, and diagonal matrices with phases
$P(\delta,1,1)={\rm diag}(e^{i\delta},1,1)$. The mixing angle measured
in atmospheric experiments is identified with $\theta_{23}$, that
measured in solar experiments with $\theta_{12}$ and that mixing angle
measured in reactor experiments, with $\theta_{13}$.

Given the matrices diagonalizing the mass matrices appearing in the
effective mass lagrangian for leptons, eq.~(\ref{effmasslag2}), and
the definition of the effective neutrino mass matrix,
eq.~(\ref{seesawfor}), such that\footnote{Note that in this convention
  $m^{\nu}_{\mrm{diag}}=L^{\nu t}_m m^{\nu}_{LL}L^{\nu}_m=L^{\nu
    \dagger}_m m^{\nu\dagger}_{LL}L^{\nu *}_m $, for
  $m^{\nu}_{\mrm{diag}}$ real.}
\be
\label{notdiagmat}
m^{l}=L^{l}_m\pmatrix{
m_{e}&0&0\cr
0&m_{\mu}&0\cr
0&0&m_{\tau}}
R^{l\dagger}_m\,,
\qquad
m_{LL}^{\nu}=L^{\nu*}_m\pmatrix{
m_{\nu_1}&0&0\cr
0&m_{\nu_2}&0\cr
0&0&m_{\nu_3}}
L^{\nu\dagger}_m\,,
\ee
then the mixing matrix $U_{MNS}$, relating mass
eigenstates $(\nu_1,\nu_2,\nu_3)$ with states participating in
neutrino oscillations, $(\nu_{e},\nu_{\mu},\nu_{\tau})$, 
\be
\pmatrix{
\nu_e\cr
\nu_\mu\cr
\nu_\tau}
=U_{\rm MNS}
\pmatrix{
\nu_1\cr
\nu_2\cr
\nu_3},
\ee
may be expressed by
\be
U_{\rm MNS}=L^{l\dagger}_m L_m^{\nu}\,.
\ee
This combination ($L^{l\dagger}_m L_m^{\nu}$) can be brought to the
form of eq.~(\ref{umnsandpsigma}), with $U$ as in
eq.~(\ref{stparmixm}).

In determining the relationship between the Dirac CP violation phase
and the phases of the mass matrix, it is usually easier to work with
the hermitean matrix $H=m^{\dagger}m$ rather than $m$ itself. Of
course the mixing angles are the same whether computed from the
diagonalization of $m$ or from that of $H$. If
$m_{\diag}=L^{\dagger}_m mR_{m}$, where $m_{\diag}$ is a real diagonal
matrix, then the hermitean matrix $H$ is diagonalized by
$m^2=L^{\dagger}HL$, where $L_{m}$ and $L$ are related through a
diagonal matrix of phases (see appendix~\ref{appA}).  Here we employ a
diagonalization of $H$ so that it may be applied for both canonical
and inverted hierarchies, although in some cases the computation of
the mixing angles from the hermitean matrix $H$ is more tedious and
obscures the simplicity of the relationship of the phases. In
appendix~\ref{appA} we detail the procedure of diagonalization; here
we just note that since $R$ and $L^{\dagger}$ are unitary matrices,
they may also be parameterized in terms of three angles and six
phases. Of these phases, only three: $\gamma^{f\prime}_{12}$,
$\gamma^{f\prime}_{13}$ and $\gamma^{f}_{23}$; can be fixed by the
elements of $H$ and the others, $\alpha^{\nu}_i$, are used to fix the
eigenvalues of $m^{\nu}_{LL}$ to be real. When constructing the
$U_{\rm MNS}$ matrix from $L^{l\dagger}$ and $L^{\nu}$ the three
undetermined phases, $a^l_i$, in $L^l$ can be used to fix the three
physical phases appearing there.  Thus we can write the
diagonalization matrices as follows
\bea
\label{paramLf}
L^f_m&=&
\!\pmatrix{
1&0&0\cr
0&e^{i(\gamma^{f\prime}_{13})}&0\cr
0&0&e^{i(\gamma^f_{23}+\gamma^{f\prime}_{13})}}\!
R^f_{23}
\!\pmatrix{
1&0&0\cr
0&e^{i(\gamma^{f\prime}_{12}- \gamma^{f\prime}_{13})}&0\cr
0&0&1}\!
R^f_{13}R^f_{12}
\!\pmatrix{
e^{-i\alpha_0}&0&0\cr
0&e^{-i\alpha_1}&0\cr
0&0&e^{-i\alpha_2}}
\nonumber\\
&=&P^f_1R^f_{23}P^f_2R^f_{13}R^f_{12}P^f_3\,.
\eea
The super-script here refers to $f=l$, for charged leptons and to
$f=\nu$, for neutrinos.  In terms of the matrices in
eq.~(\ref{paramLf}), the mixing matrix $U_{\rm MNS}$ acquires the form
\be
U_{{\rm MNS}}=P_3^{l\dagger} R^{lt}_{12}R^{lt}_{13}P^{l\dagger}_2R^{f}_{23} 
P^{l\dagger}_1P^{\nu}_1 R^{\nu}_{23}P^{\nu}_2 R^{\nu}_{13}R^{\nu}_{12}
P^{\nu}_3\,,
\ee
which we can express in terms of three angles and three physical
phases by identifying each entry of $U_{\rm MNS}$ with the parameters
appearing in the standard parameterization, eq.~(\ref{stparmixm}). A
similar construction to this has been carried out
in~\cite{King:2002nf}, we obtain:
\bea
\label{anglesofUmnsnice}
s_{13}e^{-i\delta_{O}}&=&s_{13}^{\nu}c^l_{13}c^l_{12}e^{-i\delta_1}
-c_{13}^{\nu}(s_{13}^lc^{L}_{23}c^l_{12}e^{-i\delta_2}- s_{12}^l 
s^{L}_{23} e^{-i\delta_3})=(U_{{\rm MNS}})_{e3}
\label{Umns13}
\nonumber\\
s_{12}c_{13}&=&|s_{12}^{\nu}(c_{13}^{\nu}c_{13}^lc_{12}^l
+s_{13}^{\nu}e^{i\delta_1}(c_{12}^lc_{23}^{L}s_{13}^le^{-i\delta_2}
+s_{12}^ls_{23}^{L}))
+c^{\nu}_{12}s^l_{13}c^l_{12}s^L_{23}e^{-i\delta_2}
\nonumber\\
&&- c^{\nu}_{12}s_{12}^l c_{23}^{L}e^{-i\delta_3}|
\nonumber\\
c_{12}c_{13}&=&|c_{12}^{\nu}(c_{13}^{\nu} c_{13}^lc_{12}^l
+ s_{13}^{\nu}e^{i\delta_1}(c_{12}^lc_{23}^{L}s_{13}^le^{-i\delta_2}
+s_{12}^ls_{23}^{L}))-s_{12}^{\nu}c_{12}^ls_{13}^ls^{L}_{23}e^{-i\delta_2}
\nonumber\\ 
&& +s^{\nu}_{12}s^l_{12}c_{23}^{L}e^{-i\delta_3}|
\nonumber\\
s_{23}c_{13}&=&|s^{L}_{23}c_{12}^lc_{13}^{\nu}e^{-i\delta_3}
-c_{23}^{L}c_{13}^{\nu}s_{12}^ls_{13}^le^{-i\delta_2}
+c_{13}^ls^l_{12}s_{13}^{\nu}e^{-i\delta_1}|
\label{sines23}
\nonumber\\
c_{23}c_{13}&=&|c^{L}_{23}c_{13}^lc^{\nu}_{13}e^{-i\delta_2}
+s^l_{13}s^{\nu}_{13}e^{-i\delta_1}|\,,\quad{\rm for}
\nonumber\\ 
\theta_{23}&=&\theta_{\rm atm}\,,\quad 
\theta_{12}=\theta_{\rm sol}\,,\quad 
\theta_{13}=\theta_{\rm rct}\,,
\eea
where 
\bea
\label{phasesumns}
\delta_1 &=& \gamma^{\nu\prime}_{13}-\gamma^{\nu\prime}_{12}-(\xi_s-\xi_c)
\nonumber\\ 
\delta_3 &=&(\gamma_{12}^{l\prime}-\gamma_{12}^{\nu\prime})+(\chi-\xi_c)\,,
\qquad
\delta_2 =\gamma_{12}^{\nu\prime}+\gamma_{13}^{l\prime} 
-\gamma^{l\prime}_{12}\,,
\qquad
\chi=\gamma_{23}^{l}-\gamma_{23}^{\nu}
\nonumber\\
\xi_c&=&{\rm Arg}(s^l_{23}s_{23}^{\nu}+c^l_{23}c^{\nu}_{23}e^{-i\chi})\,,
\qquad
\xi_s={\rm Arg}(-s^l_{23}c_{23}^{\nu}+c^l_{23}s^{\nu}_{23}e^{-i\chi})
\nonumber\\
c^{L}_{23}&=&|s_{23}^{\nu}s_{23}^l+c_{23}^{\nu}c_{23}^le^{i\chi}|\,,
\qquad
s^{L}_{23}=|c^l_{23}s^{\nu}_{23}e^{-i\chi}-s^l_{23}c^{\nu}_{23}|\,.
\label{chi} 
\eea
Using these formulas one can readily identify the following
interesting points: $(a)$ When we are working in the flavour basis, we
can see that the Dirac CP violation phase, $\delta_{\rm O}$ is given
by $\delta_1=\gamma^{\nu\prime}_{13}-\gamma^{\nu\prime}_{12}$, as we
have noted previously. $(b)$ The mixing angle $\theta_{13}$ has
contributions from $\theta_{12}^l$, $\theta_{13}^{\nu}$ and
$\theta_{13}^{l}$ such that if these two last terms are negligible in
comparison to the first one, we can relate the reactor angle to
parameters of the charged lepton mass matrix~\cite{Ross:2002fb}. $(c)$
The angle $\theta_{23}$ has a term that goes like
$\theta_{23}^{\nu}-\theta^l_{23}$, so even if the effective neutrino
mass matrix in the symmetry basis is such that $t^{\nu}_{23}=1$, there
will be a deviation from maximality due to the charged leptons. The
same happens in the case of $\theta_{12}$, this is particularly
relevant for models that predict $\theta^{\nu}_{12}$ nearly maximal
because the $\theta^l_{12}$ contribution can bring it down to the
appropriate value, eq.~(\ref{expcons}). $(d)$ There are some models that
predict $\theta^{\nu}_{13}$ to small or too big in comparison to the
limit of eq.~(\ref{expcons}), however an appropriate contribution from
$s^l_{13}$ or $s^l_{12}$ could agree with the limit.

The strength of CP violation in neutrino oscillations is expressed in
terms of the invariant ${\rm Im}[(U_{\rm MNS})_{11}(U_{\rm
    MNS})_{22}(U_{\rm MNS})^*_{12}(U_{\rm MNS})^*_{21}]$ which can be
written as
\be
J_{\rm CP}=\frac{1}{8}\sin(2\theta_{12})\sin(2\theta_{13})
\sin(2\theta_{23})\sin(\delta_{\rm O})
\ee
In the flavour basis we can write this invariant in terms of the
hermitean matrix $H^{\nu}=m^{\nu\dagger} m^{\nu}$ with phases
$\gamma^{\nu}_{ij}$, using eqs.~(\ref{gammapridef}), we have
\be
{\rm J}_{\rm CP}=-\frac{{\rm Im}[ (H^{\nu}_s)_{12} (H^{\nu}_s)_{23}  
(H^{\nu}_s)_{31}] }{\Delta m^2_{21} \Delta m^2_{31} \Delta m^2_{32}}\,,
\ee
which can be used without constructing the $U_{\rm MNS}$ matrix as has
been pointed out in~\cite{Branco:2002ws}.

\section{CP violation phase from  neutrino oscillations}\label{CanHier}\label{section4}

\subsection{Predictions of a class of hierarchical neutrino mass matrices}\label{predcanhier}\label{section4.1}

We assume that the low energy neutrinos acquire their mass through the
see-saw mechanism, eq.~(\ref{seesawfor}). Here, let us write
explicitly the form of the effective Majorana mass matrix
$(m^{\nu}_{LL})^{\dagger}=$ 
\be
\label{majmasseff}
\pmatrix{
\frac{m^{ D 2}_{12}}{M_2}+\frac{m^{D 2}_{13}}{M_3}& 
\frac{m^{D}_{12}m^{D}_{22}}{M_2}+\frac{m^{D}_{13}m^{D}_{23}}{M_3}&
\frac{m^{D}_{12}m^{D}_{32}}{M_2}+\frac{m^{D}_{13}m^{D}_{33}}{M_3} \cr
\frac{m ^{D}
_{21}m^{D}_{22}}{M_2}+\frac{m^{D}_{13}m^{D}_{23}}{M_3}&
\frac{m^{D 2}_{21}}{M_1}+\frac{m^{D 2}_{22}}{M_2}+\frac{m^{D 2}_{23}}{M_3}&
\frac{m^{D}_{21}m^{D}_{31} }{M_1}+\frac{m^{D}_{22}m^{D}_{32}}{M_2}+
\frac{m^{D}_{23}m^{D}_{33}}{M_3} \cr
\frac{m^{D}_{12}m_{32}}{M_2}+\frac{m^{D}_{13}m^{D}_{33}}{M_3}& 
\frac{m^{D}_{21}m^{D}_{31} }{M_1}+\frac{m^{D}_{22}m^{D}_{32}}{M_2}
+\frac{m^{D}_{23}m^{D}_{33}}{M_3}&\frac{m^{D 2}_{31}}{M_1}
+\frac{m^{D 2}_{32}}{M_2}+\frac{m^{D 2}_{33}}{M_3}
},
\ee
where we have written $m^D=m^{\nu}_{D}$ for simplicity and have
assumed that $m^{\nu}_{D11}=0$ and $m^{\nu}_{{\rm{M}}}={\rm diag}(M_1,
M_2, M_3)$.  Under the hierarchy
\be
\label{subsequdo}
\frac{|m^{\nu}_{D31}|^2,|m^{\nu}_{D21}|^2,|m^{\nu}_{D21}m^{\nu}_{D31}|}{M_1}
\gg\frac{m^{\nu}_{Di2}m^{\nu}_{Dj2}}{M_2}\gg\frac{m^{\nu}_{Di3}m^{\nu}_{Dj3}}
{M_3}\,;\qquad i,j=1,2,3\,;
\ee
which is often referred as \emph{right-handed neutrino sub-sequential
  dominance}~\cite{King:1999mb, King:2002nf} it is possible to explain
large mixing angles for atmospheric and solar neutrinos and an small
reactor neutrino mixings, see appendix~\ref{appA} for the form of the
mixing angles and masses.  An specific realization of this pattern has
been presented in~\cite{Ross:2002fb}, and we will discuss the
implications for CP violation in neutrino oscillations in the next
subsection.  At the moment let us analyze, in this class of models,
the determination of the CP violation phase in the symmetry
basis. Note from eqs.~(\ref{anglesofUmnsnice}),~(\ref{phasesumns}) that
this is given in terms of the angles $\theta^{\nu}_{13}$,
$\theta^{\nu}_{12}$ and $\theta^{\nu}_{23}$, entering in the
diagonalization of the effective neutrino mass matrix, the angles
$\theta^l_{13}$, $\theta^l_{12}$ and $\theta^l_{23}$, entering in the
diagonalization of the charged lepton matrix, and the phases
$\gamma^{f\prime}_{12}$, $\gamma^{f\prime}_{13}$ and
$\gamma^{f\prime}_{23}$ for $f=\nu,l$. Note that
$\gamma^{\nu\prime}_{12}$, $\gamma^{\nu\prime}_{13}$ can be determined
from eqs.~(\ref{gammapridef}), and $\gamma_{23}$ can be obtained from
eq.~(\ref{Gammaij}), which is equivalent to ask for a real tangent of
the mixing angle $\theta_{23}$ -eq.~(\ref{equivg23betas}),
\be
\gamma^{\nu}_{23}=\phi^D_{31}-\phi^D_{21}
\ee
and, with this information and eq.~(\ref{tanthetaij}), equivalent to
the condition of having a positive real tangent $t^{\nu}_{12}$, we
have
\bea
\label{condchida}
c^{\nu}_{23}|m^{\nu}_{D22}|\sin(\xi_{22})&=&
s^{\nu}_{23}|m^{\nu}_{D23}|\sin(\xi_{23})\,,
\nonumber\\
-\xi_{22}&=&\phi^D_{12}-\phi^D_{22}+\gamma^{\nu\prime}_{12}
\nonumber\\
-\xi_{32}&=&\phi^D_{12}-\phi^D_{32}+\gamma^{\nu}_{23}
+\gamma^{\nu\prime}_{12}\,.
\eea
Let us call
$\delta^{\nu}=\gamma^{\nu\prime}_{13}-\gamma^{\nu\prime}_{12}$, a
motivation for this definition is that in the limit of the flavour
basis this combination is the Dirac CP violation phase in the lepton
sector $\delta_{O}$, eq.~(\ref{phasesumns}). Thus $\xi_{22}$ and
$\xi_{23}$ can be rewritten as
\be
\label{xisimp}
\xi_{22}=\phi^D_{22}-\phi^D_{12}-\gamma^{\nu\prime}_{13}+\delta^{\nu}\,,
\qquad
\xi_{32}=\phi^D_{32}-\phi^D_{31}+\phi^D_{21}-\phi^D_{12}-
\gamma^{\nu\prime}_{13}+\delta^{\nu}\,.
\ee
From eq.~(\ref{gammapridef}) we can determine
$\gamma^{\nu\prime}_{13}$, to leading order, to obtain
\bea
\label{gammanupri}
\gamma^{\nu\prime}_{13}&=&\phi^D_{21}-\phi^D_{12}-\eta_{12}
\nonumber\\ 
\eta_{12}&\equiv&{\rm Arg}[m^{\nu *}_{D22}m^{\nu}_{D21}+m^{\nu *}_{D32}
m^{\nu}_{D31}]
\eea
inserting the expression for $\gamma^{\nu\prime}_{13}$ in
eqs.~(\ref{xisimp}),~(\ref{condchida}) we obtain
\be
\label{chidisima}
\tan(\eta_{12}+\delta^{\nu})\approx
\frac{|m^{\nu}_{D22}|c^{\nu}_{23}\sin(\phi^D_{22}-\phi^D_{21})
-|m^{\nu}_{D32}|s_{23}\sin(\phi^D_{32}-\phi^D_{31})}{-|m^{\nu}_{D22}
|c^{\nu}_{23}\cos(\phi^D_{22}-\phi^D_{21})+|m^{\nu}_{D32}
\cos(\phi^D_{32}-\phi^D_{31})|}\,.
\ee
If we have $|m^{\nu}_{D22}c^{\nu}_{23}|= |m^{\nu}_{D32}s^{\nu}_{23}|$ then
\bea
\label{deltanu}
\delta^{\nu}&\approx&-2\eta_{12}-\frac{\pi}{2}
\nonumber\\ 
\eta_{12}&\approx &\frac{(\phi^D_{21}-\phi^D_{22}+\phi^D_{31}-\phi^D_{32})}{2}
\,.
\eea
In the real case, as can seen from eq.~(\ref{chidisima}),
$\delta^{\nu}=0$. In the limit in which $|m^D_{22}|=0$, which could
give also maximal mixing, we have
\be
\delta^{\nu}=-2\eta_{12}=(\phi_{32}-\phi_{31})
\ee
which agrees with the result presented in~\cite{King:2002qh}.

\subsection{Predictions of a class of inverted hierarchical neutrino mass matrix}\label{InvHiers}\label{section4.2}

It is possible to obtain an inverted hierarchy $IH2$,
eq.~(\ref{plausinvhi}), describing the mass splittings and mixings of
the low energy neutrinos, under the following conditions 
\bea
\label{subsequdo2}
\frac{m^{\nu}_{Di2}m^{\nu}_{Dj2}}{M_2}&\gg&\frac{m^{\nu}_{Di3}
m^{\nu}_{Dj3}}{M_3}\,;
\quad 
i,j=1,2,3\,;
\nonumber\\ 
\frac{|m^{\nu}_{D31}|^2,|m^{\nu}_{D21}|^2,|m^{\nu}_{D21}
m^{\nu}_{D31}|}{M_1}&\gg&\frac{m^{\nu}_{Dk2}m^{\nu}_{Dl2}}{M_2}
\quad 
k,l=2,3\,;
\nonumber\\ 
\frac{m^{\nu}_{D21}m^{\nu}_{D31}}{M_1}&=&O
\left(\frac{m^{\nu2}_{12}}{M_2}\right).
\eea
Using the procedure in appendix~\ref{appA}, we can diagonalized the
matrix $m^{\nu}_{\rm LL}$, with a diagonalization matrix $L_m$ of the
form eq.~(\ref{diagmatLm}), except that now
\be
\beta_1=\gamma^{\nu}_{23}+\gamma^{\nu\prime}_{13}+\pi\,,
\qquad 
\beta_2=\gamma^{\nu\prime}_{13}\,, 
\qquad \beta_3=\gamma^{\nu\prime}_{12}-\gamma^{\nu\prime}_{13}\,.
\ee
In this case the mixing angles are given approximately by
\bea
\label{tanmixinvh}
t^{\nu}_{23}&\approx& \frac{|m^{\nu}_{D31}|}{|m^{\nu}_{D21}|}
\nonumber\\
t^{\nu}_{13}&\approx& \frac{|m^{\nu}_{D31}m^{\nu}_{D22}-m^{\nu}_{D21}
m^{\nu}_{D32}|} {|m^{\nu}_{D12}|\sqrt{|m^{\nu}_{D21}|^2+ |m^{\nu}_{D31}|^2}}
\nonumber\\
t^{\nu}_{12}&\approx& 1-\frac{\Delta_{12}}{2}
\nonumber\\ 
\Delta_{12}&=&\frac{\sqrt{|m^{\nu}_{D21}|^2+ |m^{\nu}_{D31}|^2}}{|
m^{\nu}_{D12}|}\left(\frac{\frac{M_2}{M_1}(|m^{\nu}_{D21}|^2+ 
|m^{\nu}_{D31}|^2)-|m^{\nu}_{D12}|^2}{|m^{\nu}_{D22}m^{\nu *}_{D21} 
+ m^{\nu}_{D32} m^{\nu *}_{D31}| } \right)\cos(\eta_{12})
\nonumber\\ 
\eta_{12}&\equiv &{\rm Arg}[m^{\nu *}_{D22}m^{\nu}_{D21}+m^{\nu *}_{D32}
m^{\nu}_{D31}]\,.
\eea

If the matrix $m^{\nu}_{\rm LL}$ is to be realized in the flavour
basis, then in order for $t^{\nu}_{23}$ to account for the mixing of
the atmospheric neutrinos, $m^{\nu}_{21}$ and $m^{\nu}_{31}$ need to
be of the same order. From the expression of $t^{\nu}_{13}$ in
eq.~(\ref{tanmixinvh}), we can see that in order for this mixing to
describe the reactor experiments, there should be a cancellation
between $m^{\nu}_{D22}$ and $m^{\nu}_{D32}$. Finally
$\frac{M_2}{M_1}(|m^{\nu}_{D21}|^2+|m^{\nu}_{D31}|^2)-|m^{\nu}_{D12}|^2$
is constrained to reproduce $\Delta_{12}$ small, $\Delta_{12}\approx
(0.21, 0.77)$, according to eq.~(\ref{expcons}), in order to account
for the mixing of the solar neutrino experiments.

The masses of the low energy neutrinos are given approximately by 
\bea
\label{massinvh}
m_{\nu_3}&\approx & \frac{m^{\nu 2}_{D12}}{M_1}s^{\nu 2}_{13}
\nonumber\\ 
m_{\nu_2}&\approx & e^{-2i\phi_{12}} \left( c^{\nu 2}_{12} m^{\nu\prime}_{11}
+ s^{\nu 2}_{12} e^{2i\gamma^{\prime}_{12}} m^{\nu\prime}_{22}  + 
\frac{2c^{\nu}_{12}s^{\nu}_{12}}{M_2} \frac{(m^{\nu}_{D21}m^{\nu *}_{D22}
+ m^{\nu}_{D31}m^{\nu *}_{D32})}{\sqrt{|m^{\nu}_{D21}|^2+ 
|m^{\nu}_{D31}|^2}} \right) 
\nonumber\\ 
m_{\nu_1}&\approx& e^{-2i\phi_{12}}\left(s^{\nu 2}_{12}m^{\nu\prime}_{11} 
+ c^{\nu 2}_{12} e^{2i\gamma^{\prime}_{12}} m^{\nu\prime}_{22} 
- \frac{2c^{\nu}_{12}s^{\nu}_{12}}{M_2} \frac{(m^{\nu}_{D21}m^{\nu *}_{D22}
+ m^{\nu}_{D31}m^{\nu *}_{D32})}{\sqrt{|m^{\nu}_{D21}|^2+
 |m^{\nu}_{D31}|^2}} \right),
\nonumber
\eea
where
\be
|m^{\nu\prime}_{22}|=\frac{|m^{\nu}_{D21}|^2+|m^{\nu}_{D31}|^2}{|M_1|}\,,\qquad
|m^{\nu\prime}_{11}|=\frac{\leb m^{\nu}_{D12}\rib^2}{|M_2|}\,.
\ee
Note from eq.~(\ref{subsequdo2}) that the terms in
eq.~(\ref{massinvh}) multiplying $2c^{\nu}_{12}s^{\nu}_{12}$ are
smaller than the terms multiplying $(c^{\nu}_{12})^2$ or
$(s^{\nu}_{12})^2$, and since $|m^{\nu\prime}_{22}|$ and
$|m^{\nu\prime}_{11}|$ need to be very close to each other in order to
reproduce a small $\Delta_{12}$, then $m_{\nu_1}\lesssim m_{\nu_2}$.
The phases appearing in~(\ref{massinvh}) can be computed from
eqs.~(\ref{gammapridef}) or can be determined from the conditions of
having real values for the tangents of the mixing angles. Thus we have
\bea
\label{gammasinvhi}
\gamma^{\nu}_{23}&\approx& \phi^D_{31}-\phi^D_{21}
\nonumber\\ 
\gamma^{\nu\prime}_{13}&=&(\phi^D_{21}-\phi^D_{12})-\eta_{12}\,.
\eea
We determine the phase
$\delta^{\nu}=\gamma^{\nu\prime}_{13}-\gamma^{\nu\prime}_{12}$, as in
section~\ref{predcanhier}.  From eq.~(\ref{tanthetaij}), or from the
condition of having a positive real tangent $t^{\nu}_{12}$, we have
that
\bea
\label{mnudef}
|m^{\nu\prime}_{22}|\sin(\xi_{22})&=&|m^{\nu\prime}_{11}|\sin(\xi_{11})
\nonumber\\ 
\xi_{22}&=&\phi^{D\prime}_{22}-\phi^{D\prime}_{12}\,, \qquad
\xi_{22}=\phi^{D\prime}_{11}-\phi^{D\prime}_{12}\,,
\eea
where
\bea
\label{defphispr}
\phi^{D\prime}_{22} &=& 2(\gamma^{\nu\prime}_{13}-\phi^D_{21})\,,
\qquad
\phi^{D\prime}_{11}=-2\phi^D_{12}, 
\nonumber\\ 
\phi^{D\prime}_{12}&=&\gamma^{\nu\prime}_{13}-\phi^{D}_{12}
-\phi^{D}_{21}+\eta_{12}.
\eea
Thus, inserting eq.~({~\ref{defphispr}) into eq.~(\ref{mnudef}), we
  can write for $\delta^{\nu}$ the following expression
\be
\label{chidisma2}
\tan(\delta^{\nu})=\frac{|m^{\nu\prime}_{22}|\sin(-2\eta_{12})}{|
m^{\prime}_{22}|\cos(-2\eta_{12})+|m^{\nu\prime}_{11}|}\,.
\ee
In the limit $|m^{\nu\prime}_{22}|=|m^{\nu\prime}_{11}|$,
$\delta^{\nu}$ is simply given by
\be 
\label{deltanuinvh}
\delta^{\nu}=-\eta_{12}={\rm Arg}[m^{\nu *}_{D22}m^{\nu}_{D21}
+m^{\nu *}_{D32}m^{\nu}_{D31}]\,.
\ee

\subsection{Examples}\label{examphier}\label{section4.3}

\subsubsection{Predictions for the texture inspired in a $\SU(3)_F\times \SO(10)$  symmetry}

Here we consider matrices of the form of eq.~(\ref{massmatpar}), where
we have mentioned that only the solution $\kappa =2$, $a_{l}=+3$,
$a_{\nu }=0$ produces a large atmospheric and solar mixing compatible
with experiments~\cite{Ross:2002fb}. In this case
\be
Y^{\nu}_{22}={\mathcal{O}}(Y^{\nu}_{23},Y^{\nu}_{12},Y^{\nu}_{13})
\ee
and with a diagonal matrix for right-handed neutrinos such that
$M_1\ll M_2\ll M_3$, we have then an effective Majorana mass matrix
$m^{\nu}_{LL}$, given by eq.~(\ref{seesawfor}), which satisfies the
conditions~(\ref{subsequdo}). In this case, therefore we can use the
form of the phases $\gamma^{\nu\prime}_{ij}$ obtained in
section~\ref{predcanhier}.

Since we are working in the symmetry basis we need to include the
mixings from the charged lepton sector.  The matrix~(\ref{massmatpar})
for $f=l$ can also be diagonalized by the procedure detailed in
appendix~\ref{appA}, but in this case the diagonalization process
produces small mixing angles. From eqs.~(\ref{gammapridef}) we can
determine the relevant phases contributing to the $U_{\rm MNS}$ matrix
\be
\gamma^{l\prime}_{12}\approx\gamma^l_{12}=\phi^l_{22}-\phi^l_{12}\,,
\qquad
\gamma^{l\prime}_{13}\approx \gamma^l_{13}-\gamma^l_{23}= \phi^l_{23}
-\phi^l_{13}\,.
\ee
The lepton  mixing angles are given by
\be
\label{chleptsmI}
s^l_{12}\approx\frac{|m^l_{12}+\frac{m^l_{13}m^l_{23}}{m_{33}^l}|}{|
m_{22}^l+\frac{m_{23}^{l2}}{m_{33}^l}|}={\mathcal{O}}(\bar{\epsilon})\,,
\quad 
s_{23}^l\approx \frac{|m^l_{23}+\frac{m^l_{32}m^l_{22}}{m^l_{33}}|}{
|m^l_{33}|}\,,
\quad 
s^l_{13}\approx\frac{|m^l_{13}-\frac{m^l_{12}m^l_{23}}{m_{33}^l}|}{
|m_{33}^l|}={\mathcal{O}}(\bar{\epsilon}^3)\,.
\ee
With this information and the formulas appearing in
eq.~(\ref{anglesofUmnsnice}) we can see that the the solar and
atmospheric mixing angles are mainly given by the angles
$\theta^{\nu}_{12}$ and $\theta^{\nu}_{23}$, respectively, appearing
in the diagonalization of the neutrino mass matrix~(\ref{majmasseff}):
\bea
\label{mixmalyne}
\tan\theta_{\rm sol}=|t_{12}|&\approx&|t^{\nu}_{12}|\left |\frac{1
-\frac{s^l_{12}c^{L}_{23}e^{-i\delta_3}}{t^{\nu}_{12}}}{1+s^l_{12}
t^{\nu}_{12}c^{L}_{23}e^{-i\delta_3}}\right| 
\nonumber\\
\tan\theta_{\rm atm}=|t_{23}|&\approx&|t^{L}_{23}|\left|
\frac{1-\frac{s^l_{12}s^{\nu}_{13}}{t^{L}_{23}}
e^{-i(\delta_1-\delta_3)}}{1+\frac{s^l_{13}s^{\nu}_{13}}{c^{L}_{23}}
e^{-i(\delta_1-\delta_2)}
}\right|
\nonumber\\ 
c^{L}_{23}\approx c^{\nu}_{23}|1+2 t^{\nu}_{23}\theta^l_{23}\cos\chi|\,,
&\quad&
s^{L}_{23}\approx s^{\nu}_{23}|1-2t^{\nu -1}_{23}
\theta^l_{23}\cos\chi|\,,
\eea
where the phases are given in eqs.~(\ref{phasesumns}).  Note however
that the mixing angle explained by reactor experiments, $\theta_{13}$,
can receive important contributions from $\theta^{l}_{12}$, a mixing
angle entering in the diagonalization of the charged lepton mass
matrix, eq.~(\ref{massmatpar}), thus we have
\be
\sin{\theta_{\rm rct}}e^{-i\delta_{\rm O}}=s_{13}e^{-i\delta_{\rm O}}
\approx \theta^{\nu}_{13}e^{-i\delta_{1}} +
\theta^l_{12}\sin^L_{23}e^{-i\delta_{3}}\,.
\ee
In order to determine $\delta_{\rm O}$ in this case we need to see
which are the dominant terms in eq.~(\ref{Umns13}). We consider here
three such cases, $s^{\nu}_{13}\gg s^l_{12}$, $s^{\nu}_{13}\ll
s^l_{12} $ and $s^{\nu}_{13}={\mathcal{O}}(s^l_{12})$. These cases
would correspond to \emph{different} models for which the
contributions from the charged leptons to the mixing matrix $U_{\rm
  MNS}$ are more or less important, and hence the results \emph{are
  not} equivalent.
\begin{itemize}
\item[$(a)$]$\mathbf{s^{\nu}_{13}\gg s^l_{12}}$. In this case the CP
  violation phase appearing in neutrino oscillations would be simply
  given by
\be
\delta_{\rm O}\approx \delta_3\approx \gamma_{13}^{\nu\prime}
-\gamma_{12}^{\nu\prime}=\delta^{\nu}\,,
\ee
where $\delta^{\nu}$ is determined by eq.~(\ref{chidisima}).
\item[$(b)$]$\mathbf{s^{\nu}_{13}\ll s^l_{12}}$. In this case the
  dominant term in eq.~(\ref{Umns13}) is $s^l_{12}$ and hence
\be
\label{prediction}
\delta_{\rm O}\approx \delta_3=(\gamma_{12}^{l\prime}
-\gamma_{12}^{\nu\prime})+(\chi_c-\xi)\,.
\ee
From eqs.~(\ref{chi}) we can see that for this case
\be
\chi-\xi_c\approx \theta^l_{23}\sin\chi t^{\nu}\,,
\qquad 
\gamma_{12}^{l\prime}\approx \gamma_{12}^{l} \approx \phi^l_{22}-\phi^l_{12} 
\ee
and using eqs.~(\ref{gammanupri}) we have
\be
\label{deltafort130}
\delta_{\rm O}\approx\gamma_{12}^{l}+\eta_{12}+\delta^{\nu}+(\phi^D_{12}
-\phi^D_{21}) -\theta^l_{23}\sin\chi_c t^{\nu}_{23}\,.
\ee
In general $\delta^{\nu}$ can be determined from
eq.~(\ref{chidisima}), let us take a particular case:
$|m^{\nu}_{D22}c^{\nu}_{23}| \linebreak = |m^{\nu}_{D32}s^{\nu}_{23}|$,
so we can use eq.~(\ref{deltanu}) and hence
\be
\label{cabeq}
\delta_{\rm O}\approx (\gamma_{12}^{l}-\eta_{12}) -\frac{\pi}{2}
+(\phi^D_{12}-\phi^D_{21})-\theta^l_{23}\sin\chi t^{\nu}_{23}\,.
\ee
For the particular choice of assigning a phase $\phi$ to the element
(12) of $m^{\nu}_{{\rm{D}}}$ and a phase $\phi^{\prime}$ to the
element 13 of $m^l$ we have $\delta_{\rm O}\approx -\frac{\phi}{2}
-\theta^l_{23}\sin\phi^{\prime}$, as we have seen
in~\cite{Ross:2002fb}.
\item[$(c)$]$\mathbf{s^{\nu}_{13}\approx s^l_{12}\sin^L_{23}}$. In
  this case
\be
s_{13}e^{-i\delta_{\rm O}}\approx \theta^{\nu}_{13}e^{-i\delta_{1}} 
+ \theta^l_{12}\sin^L_{23}e^{-i\delta_{3}}\,,
\ee
thus we have 
\be
\label{deltafort13at12}
\delta_{\rm O} \approx \frac{\delta_1+\delta_3}{2}
=\delta^{\nu}+\frac{\eta_{12}}{2}+\frac{\gamma^{l\prime}_{12}}{2}
+ \frac{\phi^D_{12}-\phi^D_{21}}{2}-\frac{\theta^l_{23}\sin\chi}{2}\,.
\ee
For $|m^{\nu}_{D22}c^{\nu}_{23}|=|m^{\nu}_{D32}s^{\nu}_{23}|$, using
eq.~(\ref{deltanu}),
\be
\delta_{\rm O} \approx 
-\frac{3}{2}\eta_{12}+\frac{\gamma^l_{12}}{2}+\frac{\phi^D_{12}
-\phi^D_{21}}{2}-\frac{\theta^l_{23}\sin\chi}{2}-\frac{\pi}{2}\,.
\ee
\end{itemize}
Let us comment on a particular set of values of the mixing
angles. Suppose that
\be
\label{s12ln}
s^l_{12}\approx \sqrt{\frac{m_e}{m_{\mu}}}\approx 0.07\,,
\ee
which can be obtained using them mass matrix parameterization of
eq.~(\ref{massmatpar}), for charged leptons.  For the case of the
effective matrix for neutrinos, eq.~(\ref{majmasseff}), with the
conditions of eq.~(\ref{subsequdo}), the neutrino mixing angle
$\theta^{\nu}_{13}$ is given by
\be
t^{\nu}_{13}\approx\frac{M_1}{M_2}s^{\nu}_{23}
\frac{|m^{D*}_{12}m^{D}_{22}+m^D_{13}m_{23}^{D*}|}{|m^D_{12}|^2
+|m^D_{13}|^2}
\ee
and the ratio $M_1/M_2$ is proportional to $r_{\Delta}\equiv
\sqrt{{\Delta m^2_{\rm sol}}/{\Delta m^2_{\rm atm}}}$ due to the
constraints on the masses of the neutrinos (see
appendix~\ref{appA}). For the particular realization of
eq.~(\ref{massmatpar}) with a solution reproducing LMA
angle,~\cite{Ross:2002fb}, we have $ {M_1}/{M_2}\approx
{r_{\Delta}}/{3}$.  The latest results of
KamLAND~\cite{Eguchi:2002dm} present two valid regions for $\Delta
m^2_{12}$ at $3\sigma$, as a consequence we have two allowed regions
for $r_{\Delta}$:
\bea
\Delta m^2_{12}\in [5.1, 9.7 ]\times 10^{-5}\,{\rm eV}^2&\rightarrow & 
r_{\Delta }=0.18^{-0.08}_{+0.11}
\nonumber\\ 
\Delta m^2_{12}\in [1.2, 1.9 ]\times 10^{-4}\,{\rm eV}^2&\rightarrow & 
r_{\Delta }=0.26^{-0.10}_{+0.14}.
\eea
The first region is the one that contains the best fit point (BFP) for
$\Delta m^2_{12}=6.9\times 10^{-5}{\rm eV}^2$ and, within the
$3\sigma$ region, $s^{\nu}_{13}$ can acquire values of order $10^{-2}$
which is one order of magnitude less than $s^l_{12}$,
eq.~(\ref{s12ln}). However the BFP for $r_{\Delta}=0.16$ gives a value
of $t^{\nu}_{13}=0.05$ which is close to the value of
$s^l_{12}s^{\nu}_{23}\approx 0.07$; thus the preferred solution for
the BFP of $\Delta m^2_{12}$ points out to the third of the cases
presented above, {\bf {(c)}}, and hence the prediction for the CP
violation phase would be close to~(\ref{deltafort13at12}). In this
case $\theta_{\rm rct}=O(10^{-1})$, which agrees with the latest
bounds~\cite{Lavignac:2002gf,King:2002nf}.

\paragraph{Lepton flavour violating processes constraints.}

The model presented in this section satisfies the constraints from the
lepton flavour violating processes (LFV), $\tau\rightarrow\mu\gamma$,
$\mu\rightarrow e\gamma$. The branching ratios of these LFV,
$B(\tau\rightarrow\mu\gamma)<1.1\times 10^{-6}$ and $B(\mu\rightarrow
e\gamma)< 1.2\times 10^{-11}$, depend on the Yukawa couplings for
neutrinos and the heavy right-handed Majorana masses through the
matrix~\cite{Masina:2002qh}
\be
\label{matC}
C=Y^{\nu}{\rm Ln}\left(\frac{M_X}{M_R}\right)Y^{\nu\dagger}\,,
\ee
thus the bounds on the branching ratios can be translated in terms of
bounds for the elements $C_{\tau\mu}$ and $C_{\mu e}$,
respectively. These elements depend also on the region of the spectra
for supersymmetric particles. Small values of $(C_{\tau\mu},C_{\mu
  e})\sim (10^{-1},10^{-3})$, correspond to light susy particles
($\sim 200$\,GeV) and large values, $(C_{\tau\mu},C_{\mu e})\sim
(10^{2},10^{-2})$, correspond to heavy susy particles ($\sim
800$\,GeV)~\cite{Masina:2002qh}.  In this model we have
$C_{\tau\mu}\approx 1$ and $C_{\mu e}\approx 10^{-3}$ hence this model
would be possible for a light supersymmetric
spectra. In~\cite{Blazek:2002wq} the LFV constraints have been
analyzed for natural neutrino mass hierarchies.

\paragraph{Renormalization group equations effects.}
To express the mass splittings and phases at the electro-weak scale,
$M_{\mrm{EW}}$, the evolution of the renormalization group equations
(RGE's) from the GUT scale $M_{\mrm{G}}$ down to $M_{\mrm{EW}}$ has to
be taken into account. In the context of the see-saw mechanism, due to
the decay of the right-handed Majorana states at different scales, it
is necessary to decouple these singlets at those scales and then
consider the appropriate effective theories below them. The RGE's of
the leptonic sector have the general form~\cite{Chankowski:1993tx}
\be
\label{rgesgeral}
16\pi^2\frac{d}{dt}X_i=F_{X_i}(X_i,X_j,\ldots )\,,
\ee
where $t=\ln(\mu/\mu_0)$, $\mu$ is the running scale,
$X_i\in\{Y^\nu,M_R,Y^l,\ldots \}$ and $F_{X_i}$ is the function
describing the evolution of $X_i$. To account for these RGE's effects,
one begins with the initial conditions and the coupled differential
equations, eqs.~(\ref{rgesgeral}), at $M_{\mrm{G}}$ and then evolve them
down to the scale, $M_3$, at which the heaviest R-H Majorana neutrino
decouples. At this point the appropriate RGE's for the effective
theory need to replace those considered at the GUT scale and then it
is necessary to perform appropriate matching conditions and then
continue this process until the scale of the decoupling of the
lightest R-H Majorana state, $M_1$, is reached. At this scale the
RGE's describing the evolution of the effective five dimensional
operator producing the see-saw, eq.~(\ref{seesawfor}), can be
used. This RGE has the form~\cite{Chankowski:1993tx,
  Babu:1993qv,Antusch:2001vn} $ 16\pi^2 d m^{\nu}_{LL}/{dt}$ $=\alpha
m^{\nu}_{LL}+P^T m^{\nu}_{LL} +m^{\nu}_{LL} P$ where
$P=CY^lY^{l\dagger}$, $C=1,-3/2$ in the MSSM and the
SM~\cite{Antusch:2001ck} respectively and $\alpha$ is a function of
the Yukawa and gauge couplings. In order to account in a
quantitatively accurate way for these effects the numerical evolution
of the RGE's should be used, however qualitatively and, to a
reasonable accuracy it is useful to employ analytical formulae for the
running of masses, mixing angles and phases. In our analysis we have
employed the results of~\cite{Antusch:2003kp} were the authors have
derived analytic formulas for the running of the neutrino mixing
angles $\theta^{\nu}_{i,j}$, the mass eigenvalues $m_{\nu_i}$, the CP
violating phase $\delta^{\nu}_i$ and the Majorana phases
$\sigma_{1,2}$.  We find that for the case analyzed in this section
the effects of the RGE's in the mixing angles is less than $7\%$, for
the mass values is less than $5\%$ and for the phases is also less
than $5\%$.

\subsubsection{An example of inverted hierarchy $\mathbf{IH2}$}\label{sprealinvhier}

An explicit realization of a hierarchy $IH2$, eq.~(\ref{plausinvhi}),
satisfying the conditions~(\ref{subsequdo2}), can be obtained with the
following matrices 
\be
\label{exinvhier}
Y^{\nu}=\frac{m^{\nu}_D}{|m^{\nu}_{D 33}|}=\pmatrix{
0&a_{12}\lambda_2 & a_{13}\lambda_2\cr
a_{21}\lambda_1 & a_{22}\lambda_2\epsilon & a_{32}\lambda_2\epsilon\cr
a_{31}\lambda_1 & a_{32}\lambda_2\epsilon & a_{33}\lambda_2\epsilon},
\qquad
m_{M}=\pmatrix{
M\lambda^2_1& &\cr
&M\lambda^2_2& \cr
 & & M},
\ee
where $a_{ij}$ are complex coefficients of order 1 and $\epsilon,
\lambda_{i}\ll 1$. We can see that it is possible to have
$t^{\nu}_{23}$ of order 1 since at leading order it is simply given by
$t^{\nu}_{23}=|a_{31}|/|a_{21}|$. The value of $t^{\nu}_{13}$ is
naturally small because it is given at order $\epsilon$
\be
t^{\nu}_{13}=\frac{|a_{31}a_{22} -a_{32}a_{21}|}{|a_{12}|\sqrt{|a_{21}|^2
+|a_{31}|^2}}\epsilon\,.
\ee
If this structure is to be realized in the flavour basis then to have
an acceptable value for $\Delta_{12}$, eq.~(\ref{tanmixinvh}) we need
\be
\label{ft21a31a12}
(|a_{21}|^2+ |a_{31}|^2-|a_{12}|^2)=O(\epsilon)\,,
\ee
if it is realized in the symmetry basis, then $t_{\rm sol}$ can have
contributions from the charged lepton sector~(\ref{anglesofUmnsnice})
and it is possible to relax the condition~(\ref{ft21a31a12}).  For an
inverted hierarchy we have
\bea
\Delta m^2_{\rm atm}&=&m^2_1-m^2_3\approx m^2_1\,,
\qquad {\rm for}\qquad 
|m_3|\ll |m_1|
\nonumber\\ 
\Delta m^2_{\rm sol}&=&m^2_2-m^2_1
\nonumber\\ 
|m_{\nu_2}|&=&O(10^{-2})\,{\rm eV}\,,
\eea
thus $\Delta m^2_{\rm atm}$, eq.~(\ref{expcons}), fixes the value of
$M$ to be of order $10^{15}\ {\rm GeV}$ and the value of $\Delta
m^2_{\rm sol}$ and the order of the mass of the heaviest low energy
left handed neutrino, $\nu_2$, fix the order of $\epsilon$ to be
$\lesssim {\mathcal{O}}(10^{-2})$.

The values of $\lambda_1$ and $\lambda_2$ are restricted to satisfy
LFV bounds, to this end let us write the coefficients $C_{\mu e}$ and
$C_{\tau\mu}$, described in eq.~(\ref{matC}):
\bea
C_{\mu e}&=&a_{22}a^*_{12}\lambda_2^2\epsilon \ln\left(\frac{M_X}{M\lambda^2_2}
\right)+a_{23}a^*_{13}\lambda_2^2\epsilon\ln\left(\frac{M_X}{M}\right)
\nonumber\\ 
C_{\tau\mu}&=&a_{31}a^*_{21}\lambda^2_1 \ln\left(\frac{M_X}{M\lambda^2_1}
\right)+\epsilon^2\left(\epsilon^2 a_{32}a^*_{22}\lambda^2_2 
\ln\left(\frac{M_X}{M\lambda^2_2}\right)+  a_{33}a^*_{23}\lambda^2_2 
\ln\left(\frac{M_X}{M}\right)    \right).
\eea
Here we can consider two cases: (i) $\lambda_1<\lambda_2$ and (ii)
$\lambda_1\approx\lambda_2$, both can explain the mixings observed by
neutrino oscillation experiments but would have different behaviours
for leptogenesis and could be explained by different symmetries.
\begin{itemize}
\item[(i)] In this case for the values
\be
\epsilon=4\times 10^{-3}\,,
\qquad 
\lambda_1=5\times 10^{-2}\,, 
\qquad 
\lambda_2=2.5\times 10^{-1} 
\ee
and the coefficients
\be
\label{eq:coeffIH2}
(|a_{ij}|)=\pmatrix{
0&1.84&1.2\cr
1.2&0.7&0.8\cr
1.4&1.5&1},
\qquad
(\phi_{ij})=
\pmatrix{
0&0&1.4\cr
1.8&0.2&0.4\cr
0.5&0.01&0.1},
\ee
where we have written $a_{ij}=|a_{ij}|e^{i\phi_{ij}}$,\footnote{The
  change in the coefficients does not change significatively the
  results. The only coefficient that is restricted is $a_{12}$ because
  it has to satisfy the condition of eq.~(\ref{ft21a31a12}).} we have
\bea
\label{resih2a}
\Delta m^2_{13}&=&4\times 10^{-3}\, {\rm eV}^2
\qquad
\Delta m^2_{21}=3.3\times 10^{-5}\, {\rm eV}^2
\nonumber\\ 
(t^{\nu}_{23})^2 &=& 1.36 
\qquad 
(t^{\nu}_{12})^2=0.69
\nonumber\\  
(t^{\nu}_{13})^2 &\approx & 7\times 10^{-6}
\nonumber\\ 
M_1 &=& 3.8\times 10^{12}\,{\rm GeV}
\qquad 
M_2=6.1\times 10^{13}\,{\rm GeV}
\nonumber\\ 
M &=& 1.05\times 10^{15}\,{\rm GeV}
\nonumber\\ 
m_{\nu_2}&=&0.084{\rm eV}
\qquad
m_{\nu_3}=3\times 10^{-5}{\rm eV}
\nonumber\\ 
C_{\tau\mu} &=& 0.052 
\qquad 
C_{\mu e}=0.0024\,.
\eea
Where we have given the values obtained at electro-weak scale, using
the approximate RGE's formulas of~\cite{Antusch:2003kp}. In this case,
the effects of the RGE's, for the SM or small $\tan\beta$ of the MSSM,
is an increase up to 20\% for $(t^{\nu}_{12})^2$, 1\% for
$(t^{\nu}_{23})^2$ and negligible for $(t^{\nu}_{13})^2$ and
$m_{\nu_3}$. This behaviour of $(t^{\nu}_{13})^2$ and $m_{\nu_3}$
corresponds to their smallness, compared to the other parameters,
which is in turn produced by the small value of $\epsilon\approx
0.004$. For $\Delta m^2_{21}$ there is an increase up to 60\% and for
$\Delta m^2_{32}$ up to 50\%. Larger values of $\tan\beta$ correspond
to a larger increase, if $\tan\beta\approx 20$ then there is an
increase of about 90\% in $(t^{\nu}_{12})^2$, which brings it outside
the valid experimental region to account for the solar neutrino
oscillation experiments, eq.~(\ref{expcons}). The values presented in
eq.~(\ref{resih2a}) correspond to $\tan\beta=6$. The effect of RGE's
on the CP Dirac violating phase depends on the Majorana phases, we
discuss this effect in section~\ref{sec:majpha}.

In this case we note that $\epsilon\approx\lambda_2^4$ and
$\lambda_1\approx \lambda_2^2$ so we could write the $m^{\nu}_D$ in
terms of a single parameter $\lambda=\lambda_2$, which is of the order
of the Cabibbo angle $\theta_C\approx 0.22$. In this case we have
\be 
\label{formofyukmaj}
Y^{\nu}=\pmatrix{
0&\lambda&\lambda\cr
\lambda^2&\lambda^5&\lambda^5\cr
\lambda^2&\lambda^5&y_{33}},
\qquad
M_{R}=\pmatrix{
M\lambda^4& &\cr
&M\lambda^2& \cr
 & & M},
\ee
where $y_{33}$ could be $\lambda$ or $\lambda^5$.  This symmetry can
be considered in terms of a $\U(1)_F$ symmetry for the Yukawa
couplings for leptons, in terms of eqs.~(\ref{yukawacoupne}), and for
the Majorana mass matrix, given by eq.~(\ref{rightmassu1}). Let us
consider the following assignment of charges
\be
n_{1,2,3}=2,1,0\qquad l_{1,2,3}=1,-5,-5\qquad e_{1,2,3}=11,3,5\,,
\ee
then we have
\be
Y^{\nu}=\pmatrix{
\lambda^3&\lambda^2&\lambda\cr
\lambda^3&\lambda^4&\lambda^5\cr
\lambda^3&\lambda^4&\lambda^5},
\quad 
Y^{e}=\pmatrix{
\lambda^{12}&\lambda^4&\lambda^6\cr
\lambda^4&\lambda^2&1\cr
\lambda^6&1&1},
\quad 
M_R=\pmatrix{
\lambda^{4}&\lambda^3&\lambda^2\cr
\lambda^3&\lambda^2&\lambda\cr
\lambda^2&\lambda&1}.
\ee
With appropriate coefficients for $Y^l$ it is possible to produce the
eigenvalues proportional to the masses of the charged leptons and also
small mixings for charged leptons:
\bea
\label{eigenmixchlp}
(y_e,y_{\mu},y_{\tau})&\propto& (\lambda^6,\lambda^2,1)
\nonumber\\ 
s^l_{23}&=&{\mathcal{O}}(10^{-1})\,,
\qquad 
s^l_{13}={\mathcal{O}}(\lambda^6)\,,
\qquad 
s^l_{12}={\mathcal{O}}(\lambda^2)\,.
\eea
It is also possible to reproduce $m^{\nu}_{LL}(IH2)$ of the form
eq.~(\ref{plausinvhi}). We can diagonalize $M_R$ with a matrix, $L_M$,
of small mixings such that with $(M_R)_{\rm diag}=L^t_M M_RL_M$ and
then we have
\be
Y^{\nu\prime}=Y^{\nu}L_M=\pmatrix{
a^{\prime}_{1,1}\lambda^3&a^{\prime}_{1,2}\lambda^2& a^{\prime}_{1,3}\lambda
\cr
a^{\prime}_{2,1}\lambda^3&a^{\prime}_{2,2}\lambda^4&a^{\prime}_{2,3}\lambda^5
\cr
a^{\prime}_{3,1}\lambda^3&a^{\prime}_{3,2}\lambda^4&a^{\prime}_{3,3}\lambda^5
},\qquad
(M_R)_{\rm diag}\approx\pmatrix{
\lambda^4& &\cr
&\lambda^2& \cr
& & 1}
\langle H_{\rm M} \rangle\,,
\ee
which we can identify with eqs.~(\ref{formofyukmaj}), although we
would need a cancellation for the element $a^{\prime}_{11}$ and the
coefficients would need to reproduce the order of the power of
$\lambda$ as in eq.~(\ref{formofyukmaj}).

In this case, the atmospheric and mixing angles will be dominated by
the mixings coming from the effective Majorana mass matrix, as in
eq.~(\ref{mixmalyne}), since $s^l_{23}$ and $s^l_{12}$ are small,
eq.~(\ref{eigenmixchlp}), but as we can observe from
eq.~(\ref{anglesofUmnsnice}), the reactor angle would be driven by
$s^l_{12}={\mathcal{O}}(\lambda^2)$, since
$s^l_{13}={\mathcal{O}}(\lambda^6)$ and
$s^{\nu}_{13}={\mathcal{O}}(\lambda^4)$. Thus we have
\be
\label{deltaosymmbas1}
\sin{\theta_{\rm rct}}e^{-i\delta_{\rm O}}\approx
\theta^l_{12}\sin^L_{23}e^{-i\delta_{3}}\,,\qquad 
\delta_{\rm O}= \delta_3\approx \gamma^{l\prime}_{12}
-\gamma^{\nu\prime}_{12}\,,
\ee
where $\gamma^{\nu\prime}_{12}$ and $\gamma^{\nu\prime}_{13}$ are
given by eqs.~(\ref{gammasinvhi}) and thus
\be
\delta_{\rm O}\approx \gamma^{l\prime}_{12}+(\delta^{\nu}+\eta_{12})
+(\phi^D_{21}-\phi^D_{12})\,,
\ee
and for the case of $|m^{\nu\prime}_{22}|=|m^{\nu\prime}_{11}|$, we have
\be
\label{deltaosymmbas2}
\delta_{\rm O}\approx \gamma^{l\prime}_{12}+(\phi^D_{21}-\phi^D_{12})\,,
\ee
where we have used eq.~(\ref{deltanuinvh}).
\item[(ii)] In this case for the values
\be
\epsilon=3\times 10^{-2}\,,
\qquad 
\lambda_1=1.2\times 10^{-1}\,, 
\qquad 
\lambda_2=1.4\times 10^{-1} 
\ee
and the coefficients as in eq.~(\ref{eq:coeffIH2}), we have 
\bea
\Delta m^2_{13} &=&  1.3\times 10^{-3} {\rm eV}^2
\qquad
\Delta m^2_{21}=8.4\times 10^{-5} {\rm eV}^2
\nonumber\\ 
(t^{\nu}_{23})^2 &=& 1.35
\qquad
(t^{\nu}_{12})^2=0.71
\qquad
(t^{\nu}_{13})^2\approx 2\times 10^{-4}
\nonumber\\ 
M_1 &=& 3.7\times 10^{13}{\rm GeV}
\qquad
M_2=5\times 10^{13}{\rm GeV}
\qquad
M=2.5\times 10^{15}{\rm GeV}
\nonumber\\ 
m_{\nu_2}&=&0.038{\rm eV}
\qquad 
m_{\nu_3}=0.003{\rm eV}
\nonumber\\ 
C_{\tau\mu}&=& 0.15
\qquad 
 C_{\mu e}=0.006\,.
\eea
The effects of the RGE's in the parameters at $M_{\mrm{EW}}$ for the
SM or small $\tan\beta$ of the MSSM, is similar to the previous case,
except that the increase of $(t^{\nu}_{12})^2$ is more moderate, up to
15\%. In this case due to the tendency for $\Delta m^2_{12}$ to lie in
the upper part of the allowed experimental region, see
eq.~(\ref{expcons}), large $\tan\beta$ values ($\geq 20$) bring
$\Delta m^2_{12}$ up to ${\mathcal{O}}(10^{-4}){\mrm{eV}}^2$, outside
the valid experimental region to account for solar neutrino
experiments.  For this example we note that $\lambda^2_1 \approx
2\epsilon$ then we would have
\be
\label{mat2inv}
Y^{\nu}=\pmatrix{
0&\lambda_2&\lambda_2\cr
\lambda_1&4\lambda^3_1&4\lambda^3_1\cr
\lambda_1&4\lambda^3_1&y_{33}}
\ee
where $y_{33}$ could be $\lambda_2$ or $4\lambda^4_1$. It is difficult
to motivate this pattern in the context of $\U(1)_F$ symmetries,
firstly because it requires two parameters and secondly because of the
powers appearing in each entry.
\end{itemize}

\section{CP violation in leptogenesis and its connection to neutrino oscillations}\label{Lepto}\label{section5}

\subsection{CP asymmetries from leptogenesis}\label{section5.1}

According to cosmic microwave background radiation measurements, the
observed abundance of the light elements synthesized during the Big
Bang nucleosynthesis requires that the baryon asymmetry of the
universe (BAU), parameterized by the baryon-to-entropy ratio,
$Y_B=n_B/s$, lies in the range~\cite{Olive:2002qg}
\be
\label{ybmeas}
Y_B\in(0.7, 1) \times 10^{-10}\,.
\ee
In the leptogenesis scenario~\cite{Fukugita:1986hr}, a $B-L$ asymmetry
is produced from the decay of the heavy right-handed Majorana
neutrinos, $N_j$, which violate the lepton number at a large scale
beyond the electroweak scale. The initially produced lepton asymmetry,
$Y_L$, is converted into a net baryon asymmetry $Y_B$, through the
$(B+L)$-violating sphaleron processes, such that at the end of the
processes, $Y_B$ and $Y_L$ are related by
\be
Y_B=\frac{\alpha}{\alpha-1}Y_L\,,
\qquad 
\alpha=\frac{8N_F+4N_H}{22N_F+13N_H}\,,
\ee
where $N_F$ is the number of families of heavy right handed neutrinos
and $N_H$ the number of Higgs multiplets.  In thermal leptogenesis the
right-handed neutrino number densities $Y_{N_j}$ and the generated
lepton asymmetry $Y_L$ evolve with time according to a set of
Boltzmann equations which depend on the physical processes occurring
in the thermal bath and on the expansion of the universe. Here we
assume the standard hot big bang universe, which is equivalent to
assume a very high reheating temperature after inflation, larger than
the right-handed neutrino masses, $M_j$.  In the MSSM extended with
heavy right-handed neutrinos the physical processes relevant to the
generation of BAU are typically the decays and inverse decays of $N_i$
and its scalar partners, $\widetilde{N^c}_i$, and $L$ violating
processes mediated by virtual $N_i$ or $\widetilde{N^c}_i$ particles.
Right handed neutrinos, $N_i$, decay into Higgs bosons and leptons or
into Higgsinos and s-leptons; while $\widetilde{N^c}_i$ decay into
Higgs bosons and s-leptons or into Higgsinos and leptons. In the SM,
the correspondent physical processes take place. The $CP$ asymmetries
in the different decay channels of $N_j$ and $\widetilde{N^c}_j$ can
all be expressed by the same $CP$ violation parameter
$\epsilon_j$~\cite{Plumacher:1998ru},
\bea
\label{epsj}
\varepsilon_j& =&
\frac{\Gamma (N_{R_j}\rightarrow l H_2)-\Gamma (N^{\dagger}_{R_j}\rightarrow 
l^{\dagger} H^{\dagger}_2)}{\Gamma (N_{R_j}\rightarrow l H_2)
+\Gamma (N^{\dagger}_{R_j}\rightarrow l^{\dagger} H^{\dagger}_2)} 
\nonumber\\ 
\varepsilon_j &\approx& -\frac{1}{8\pi(m^{\nu\dagger}_{fD} 
m^\nu_{fD})_{11} v^2}\sum_{i\neq j}{\rm Im}\left [(m^{\nu\dagger}_{fD} 
m^\nu_{fD})_{ji}^2\right]f\left(\frac{M^2_i}{M^2_j}\right),
\eea
where $f(x)=\sqrt{x}\leb(1+x)\ln({x}/{1+x})+(2-x)/(1-x)\rib$,
$m^\nu_D$ is the Dirac matrix for neutrinos,
eq.~(\ref{effectmasslang}), and $v_2$ is the vacuum expectation of the
Higgs field, $H_2$.  The CP asymmetries, $\epsilon_j$, are constrained
to reproduce the observed value of $Y_B$, eq.~(\ref{ybmeas}).

In the case of hierarchical heavy neutrinos, the sign of $Y_B$ is
fixed by the sign of the CP asymmetry generated in the decay of the
lightest heavy neutrino, $\varepsilon_1$. Such that for $Y_B$ to be
positive, as required by the observations, it is necessary to have
$\varepsilon_1< 0$.  Given the present measurements of neutrino masses
and oscillations, it appears plausible to associate the baryon number
of the universe with the violation of lepton number. In this context
it makes sense to determine if there is a correlation between the sign
of the baryon number of the universe and the strength of CP violation
in neutrino oscillations~\cite{Frampton:2002qc}. This correlation is
relevant to look for in successful schemes that explain or accommodate
the correct values of masses and mixings.  As we can see
from~(\ref{epsj}) this correlation will depend as well in which range
we consider for the values $M_i/M_j$. We consider here the case
$M_1<M_2\ll M_3$.  In this case, there are simplifications in the
treatment of the terms that enter in $\epsilon_j$. For $M_1< M_2\ll
M_3$, $f(x)\approx -\frac{3}{2\sqrt{x}}$, $x\gtrsim
15$~\cite{Buchmuller:1998yu,Branco:2002kt,King:2002nf}, then the only
relevant $CP$ asymmetry is the one produced by the lightest right
handed neutrino, which can be expressed by
\be
\label{eps1}
\varepsilon_1\approx -\frac{3}{16\pi(m^{\nu\dagger}_{fD} 
m^\nu_{fD})_{11}v^2}\sum_{i\neq 1}{\rm Im}
\left [(m^{\nu\dagger}_{fD} m^\nu_{fD})_{1i}^2\right]\frac{M_1}{M_i}\,,
\ee
this expression is given in the basis where both the charged leptons
and the heavy right-handed neutrinos are diagonal.

In the context of thermal leptogenesis, when the observed baryon
asymmetry is generated through the decays of the lightest heavy
Majorana neutrino $N_1$, in order to produce the right baryon
asymmetry, $Y_B$, there exists an upper bound on the lightest $M_1$,
typically $M_1\gtrsim 10^{8}{\rm GeV}$~\cite{Davidson:2002qv}. Thermal
leptogenesis requires that the reheating temperature $T_R$ be such
that $M_1\lesssim T_R$, in this context a model independent
analysis~\cite{Buchmuller:2002rq} has given a constraint of the
re-heating temperature to be $T_R\approx M_1=
{\mathcal{O}}(10^{10}{\rm GeV})$. This temperature is marginally
compatible with the maximum allowed one in supergravity theories,
usually $T_R\lesssim (10^8-10^9) {\rm GeV}$, which is usually
constrained by thermal gravitino production.  Thus we have a slight
incompatibility between the reheating temperature $T_R$ required by
thermal leptogenesis and the one required by many supergravity
theories.  There are many options to overcome this problem. For
example, it is possible to consider, still in the context of thermal
leptogenesis, the decays of two heavy neutrinos which are
quasi-degenerated in mass $M_1\approx M_2$,~\cite{Branco:2002kt}. In
this case, the $CP$ asymmetries $\varepsilon_j$ are enhanced due to
self-energy contributions and the required baryon asymmetry can be
produced by right-handed heavy neutrinos with masses $M_1\approx
M_2\lesssim 10^{8}\ {\rm GeV}$ and reheating temperatures, $T_R$, of
that order.  Other options to lower $T_R$, and which have more freedom
about $M_1$, include non-thermal production
mechanisms~\cite{Giudice:1999fb,Mazumdar:1999tk} where the condition
$M_1\lesssim T_R$ is not required, gravitationally suppressed decay of
the inflaton in models of high scale inflation~\cite{Ross:1996dq} and
low scale inflationary models~\cite{Randall:1996dj, Mazumdar:1999tk,
  German:2001tz}.

\subsection{Estimation of $Y_B$}\label{section5.2}

In order to evaluate the baryon asymmetry, $Y_B$, we need $Y_L$, which
is given by
\be
Y_L=d_L\frac{\varepsilon_1}{g^*}\,,\qquad d_L=(1-\alpha)d_{B-L}\,,
\ee
where $g*$ is the effective number of degrees of freedom, for the SM
$g^*=106.75$ and for the MSSM, $g^*=228.75$. The parameter $d$ is the
\emph{dilution factor}, which takes into account the washout effects
produced by inverse decays and lepton number violating
scatterings. For different models we should integrate numerically the
set of Boltzmann equations for the lepton asymmetry $Y_L$ and the
asymmetries produced by the right handed Majorana neutrinos,
$Y_{N_j}$. In our case we would like to give just an approximation of
$Y_L$, in order to see whether a possible hierarchy can be realized
within the thermal leptogenesis scenario, thus we use the
approximation obtained in~\cite{Hirsch:2001dg}.  There, it has been
obtained an \emph{empirical} formula for the value of
$Log_{10}(d_{B-L})$, which is taken to be the smallest of the
following quantities
\bea
Log_{10}(d_{B-L})&=&0.8 Log_{10}(\tilde{m}_1)+1.7+0.05 Log_{10}(M^{10}_1)
\nonumber\\ 
Log_{10}(d_{B-L})&=& -1.2-0.05  Log_{10}(M^{10}_1)
\nonumber\\ 
Log_{10}(d_{B-L})&=& -(3.8+Log_{10}(M^{10}_1)(Log_{10}(\tilde{m}_1)+2)-
\nonumber\\ 
&&-\left(5.4-\frac{2}{3} Log_{10}(M^{10}_1)\right)^2 -\frac{3}{2}\,,
\eea
where 
\be
\tilde{m}_1=\frac{\left(m^{\nu\dagger}_{D}m^{\nu}_{D}\right)_{11}}{M_1}\,,
\qquad
M^{10}_1=\frac{M_1}{10^{10}\,{\rm GeV}}\,.
\ee

\subsection{Relative sign between $\varepsilon_j$ and $J_{CP}$ and relation of phases}\label{section5.3}

Now we can study the correlation between the sign of $\varepsilon_j$
and the sign of $J_{CP}$, for the different models presented in
section~\ref{CanHier}. These possible correlations would be satisfied
at the scale of the decay of the lightest R-H neutrino. After this the
quantities appearing in the CP leptogenesis asymmetries,
eq.~(\ref{epsj}), would evolve differently~\cite{Antusch:2003kp}.

\paragraph{I.}
For the models presented in section~\ref{examphier}, we have that
${\mathbf{M_1< M_2\ll M_3}}$ so we can use eq.~(\ref{eps1}) to
evaluate $\varepsilon_1$, but we need to translate it to the symmetry
basis, however, in the quantity
$H^D=m^{\nu\dagger}_{fD}m^{\nu}_{fD}$\footnote{Here the index $f$
  corresponds to quantities in the flavour basis and $s$ to quantities
  in the symmetry basis and $H^D_b=m^{\nu\dagger}_{Db} m^{\nu}_{Db}$
  for $b=f,s$.} the contribution from the rotation to the base in
which the charged leptons are diagonal cancels, thus we have
$H^D_{f}=H^D_{s}$, and hence:
\bea
\label{epsexph}
\varepsilon_1&\approx&\frac{3}{16\pi}\frac{|(H^D_{s})_{ij}|^2}{v^2|
H^D_{11}|}\frac{M_1}{M_2} \sin(\delta_{\rm L})
\nonumber\\ 
\delta_{\rm L}&=&2(\gamma^D_{12})
\nonumber\\ 
-\gamma^{D}_{12}&\equiv&{\rm Arg}[m^{\nu*}_{D21}m^{\nu}_{D22}
+m^{\nu*}_{D31}m^{\nu}_{D32}]\,,
\eea
from now on we drop the index $s$.  As we can see from
eqs.~(\ref{deltafort130}),~(\ref{deltafort13at12}), the phase
$\gamma^{D}_{12}$ is no other than $-\eta_{12}$, then
\be
\label{deltalinteta}
\delta_{\rm L}=-2\eta_{12}\,.
\ee
The phase $\eta_{12}$ enters into the expression for the CP violation
phase associated to neutrino oscillations, $\delta_{\rm O}$, for the
different cases presented in example {\bf I} of
section~\ref{CanHier}. Thus in this cases $\delta_{\rm O}$ is related
to the phase relevant for leptogenesis, $\delta_{\rm L}$. The exact
relation depends on the contribution of the elements diagonalizing the
charge lepton mixing matrix, but we will determine whether or not
there is a general feature about their relative signs.
\begin{itemize}
\item[$(a)$]$\mathbf{s^{\nu}_{13}\gg s^l_{12}}$. For this case we are
  considering effectively that $s^l_{12}=0$ and hence
  $\gamma^{l\prime}_{12}=0$, thus we have
\be
\delta_{\rm L}=-2\eta_{12}\,.
\ee
For the case
$|m^{\nu}_{D22}c^{\nu}_{23}|=|m^{\nu}_{D32}s^{\nu}_{23}|$\footnote{Note
  that in the case of $m^{\nu}_{D22}$ or $m^{\nu}_{D32}$ equal to zero
  then $\delta_{\rm O}=-2\eta_{12}$ and hence $\delta_{\rm
    L}=\delta_{\rm O}$, thus ${\rm sign} (\varepsilon_j)={\rm sign}
  (J_{CP})$.}  we have
\be
\delta_{\rm L}=\delta_{\rm O}-\frac{\pi}{2}\rightarrow {\rm sign}
(\varepsilon_1)=-{\rm sign}(J_{\rm CP})
\ee
\item[$(b)$]$\mathbf{s^{\nu}_{13}\ll s^l_{12} }$. In this case we
  compare $\delta_L$ to eq.~(\ref{cabeq}), for
  $|m^{\nu}_{D22}c^{\nu}_{23}|=|m^{\nu}_{D32}s^{\nu}_{23}|$ then we
  have
\be
\label{case01dldo}
\delta_{\rm L}\approx 2(\delta_{\rm O}-\gamma^l_{12})
+\pi-2(\phi^D_{12}-\phi^D_{21})+2\theta^l_{23}\sin\chi t^{\nu}_{23}\,.
\ee
In this case we cannot determine the sign unless for symmetric
(anti-symmetric) $m^{\nu}_D$ matrices, in which case we have that
$\phi_{12}-\phi_{21}=0 (\pi)$, then
\be
{\rm sign}(\varepsilon_1)=-{\rm sign}(J_{\rm CP})
\ee
\item[$(c)$] $\mathbf{s^{\nu}_{13}\approx s^l_{12}\sin^L_{23}}$.  In
  this case we have for
  $|m^{\nu}_{D22}c^{\nu}_{23}|=|m^{\nu}_{D32}s^{\nu}_{23}|$
\be
\label{case2dldo}
\delta_{\rm L}\approx \frac{4}{3}\delta_{\rm O}-\frac{2}{3}
\gamma^{l\prime}_{12}+\frac{2\pi}{3}-\frac{2\theta^l_{23}\sin
\chi t^{\nu}_{23}}{3}-\frac{2(\phi^D_{12}-\phi^D_{21})}{3}\,, 
\ee
where we cannot determine the sign unless we specify
$\gamma^{l\prime}_{12}$. We remark that these cases are not equivalent
since they differ in the way the charged lepton mixing angles and the
neutrino mixing angles contribute to the $U_{\rm MNS}$ matrix
elements, eq.~(\ref{anglesofUmnsnice}).
\end{itemize}

We note that for $\epsilon\approx 0.06$, as was the case presented
in~\cite{Ross:2002fb}, then $M_1\approx 10^8$\,GeV, in this case the
produced baryon asymmetry, $Y_B$, is of the order $10^{-14}$, which is
too small in comparison to the observed values, eq.~(\ref{ybmeas}).
Although this particular realization would not be valid for the
thermal leptogenesis scenario considered here,\footnote{We may also
  think in considering other options for which $M_1\approx 10^8$ can
  still be compatible with leptogenesis, such as the one mentioned
  previously within the context of thermal
  leptogenesis~\cite{Branco:2002xf}, where there are two
  quasi-degenerate right handed neutrinos, or we can consider
  non-thermal leptogenesis scenarios.~\cite{Giudice:1999fb,
    Allahverdi:2002gz}.} models based on the same structure for
masses, eq.~(\ref{massmatpar}), with $\epsilon\approx 0.2$ and
$M_1\approx 4\times 10^{10}$\,GeV, can produce a baryon asymmetry of
the correct order, eq.~(\ref{ybmeas}). In~\cite{Akhmedov:2003dg} the
authors have considered a Yukawa neutrino coupling structure similar
to eq.~(\ref{massmatpar}). They have found that is very hard to obtain
successful baryogenesis through leptogenesis because in general $Y_B$
is too small.  In~\cite{Mohapatra:1992pk} it was explored the
possibility of obtaining the correct amount of baryon asymmetry in the
universe in certain types of left-right symmetric theories with a low
right-handed scale. It turns out that the Higgs sector of the theory
should include one more real scalar field with appropriate
self-couplings. In~\cite{Falcone:2000ib,Falcone:2003af} it was
explored the possibility of achieving the correct amount of baryon
asymmetry in a model with a Yukawa neutrino matrix with a hierarchy
for the neutrino Dirac mass matrix with zeroes in the $(1,1)$ and
$(1,3)$ positions and similar entries $(2,2)$ and $(2,3)$, such that
$Y^{\nu}_{12}\ll Y^{\nu}_{2,2}\ll Y^{\nu}_{3,3}\ll 1$. The conditions
on $M_R$, in order to agree with the correct amount of baryogenesis,
were determined there.

\paragraph{II.}
For the case (i) presented in section~\ref{sprealinvhier} we have
$M_1\ll M_2\ll M_3$ thus we can use eq.~(\ref{eps1}). For the values
$\lambda\approx 0.25$ and $M_1\approx 10^{13}$\,GeV it is possible to
produce a baryon asymmetry of the order $Y_B\sim 4\times 10^{-11}$,
which is still compatible with eq.~(\ref{ybmeas}).

Let us assume first that the matrices of eq.~(\ref{exinvhier}) are
realized in the flavour basis, the relevant phase for leptogenesis is
given by $\delta_{\rm L}=-2\eta_{12}$, eq.~(\ref{deltalinteta}). If we
have that $|m^{\nu\prime}_{22}|=|m^{\nu\prime}_{11}|$ then, using
eq.~(\ref{deltanuinvh}), $\delta_{\rm L}$ and the phase appearing in
neutrino oscillations, $\delta_{\rm O}$, are simply related by
\be
\delta_{\rm L}=2\delta_{\rm O}\rightarrow {\rm sign}
(\varepsilon_1)={\rm sign }(J_{\rm CP})\,.
\ee
In the symmetry basis we need to take into account the contribution of
phases appearing in the diagonalization of charged leptons,
eq.~(\ref{deltaosymmbas1}). From eqs.~(\ref{deltaosymmbas1}) we see
that the only common phase in $\delta_{\rm O}$ and $\delta_{\rm L}$ is
$\gamma^{l\prime}_{12}$ and we cannot determine the phase unless
$\eta_{12}=0$, and we would have ${\rm sign}(\varepsilon_{1})={\rm
  sign}(J_{\rm CP})$.  As we can see for the cases $(a)$, $(b)$,
$(c)$, considered here the relation of the signs is not identified
unless there are further assumptions and/or the phases coming from
charged leptons are identified.

\section{Majorana phases and neutrinoless double beta decay}\label{sec:majpha}\label{section6}

The order of magnitude of the masses of neutrinos and the value of
Majorana phases, cannot be determined from neutrino oscillations, the
processes from which these parameters can be determined are
neutrinoless double beta decay, ($\beta\beta_{\nu 0}$ ), and tritium
beta decay. The Majorana mass term $n_L^T Cm^{\nu}_{\rm LL}n_L$
induces a $\beta\beta_{\nu 0}$ decay ($n\ n\rightarrow p\ p\ e^-_L\
e^-_L$) whose amplitude depends on the average neutrino mass
\be
\langle m_{\beta\beta} \rangle=\leb\sum^{3}_{i=1}(U^2_{\rm MNS})_{ei}
m_{\nu_i}\rib\,.
\ee
The Heidelberg-Moscow experiment~\cite{Klapdor-Kleingrothaus:2001ke}
quotes the range $\langle m_{\beta\beta} \rangle\in(0.11,0.56)\ {\rm
  eV}$ at 95\% confidence level, a result that has been widely
criticized and is hoped to be improved, given that neutrino
oscillation experiments have sensitivities of order $0.05\ {\rm
  eV}$. The improvement in this measurement it is not expected in the
near future, nevertheless, given a possible structure for neutrino
mixings it is relevant to obtain its predictions or constraints for
the Majorana phases.  In the notation of
eqs.~(\ref{convmajph}),~(\ref{umnsandpsigma}),~(\ref{stparmixm}), we
have
\bea
\label{bbnu0amp}
\langle m_{\beta\beta} \rangle&=&\leb (c_{13}c_{12})^2 e^{2i\sigma_1}m_{\nu_1}+
(c_{13}s_{12})^2 m_{\nu_2}+ (s_{13})^2 e^{2i(\sigma_2-\delta_{\rm O})}m_{\nu_3}
\rib
\nonumber\\ 
&=&\leb (c_{13}c_{12})^2 e^{i(\phi_1-\phi_2)}m_{\nu_1}+
(c_{13}s_{12})^2 m_{\nu_2}+ (s_{13})^2 e^{i(\phi_3-\phi_2)}m_{\nu_3}
\rib
\eea
so we can analyze the connections for the different cases presented in
section~\ref{CanHier}.

\paragraph{I.} 
In the limit of the strong hierarchy of eq.~(\ref{subsequdo}) we have
$\phi_1 \approx \phi_2$, as we can see from eq.~(\ref{bbnu0amp}),
nence there is only one relevant phase for $\beta\beta_{\nu 0}$,
namely
\be
\label{2bcaseI}
|\phi_{\beta\beta}|=|2(\sigma_2-\delta_{\rm O})|=|\phi_3-\phi_2|\,.
\ee
For the case considered in section~\ref{CanHier}, $\phi_3$ and
$\phi_2$ are given by
\be
\label{phasesi}
\phi_2\approx 2\phi^D_{12}\,,\qquad
\phi_3\approx -2\gamma^{\nu\prime}_{13}+2\phi^D_{21}\,.
\ee
Inserting eq.~(\ref{gammanupri}) into eq.~(\ref{phasesi}) and
substituting in eq.~(\ref{2bcaseI}) we have
\be
\label{dbdih2fv}
|\phi_{\beta\beta}|\approx |2\eta_{12}|\,.
\ee
As we can see from eq.~(\ref{deltalinteta}) in the flavour basis, the
phase appearing in the amplitude of neutrinoless double beta decay and
the phase relevant for leptogenesis are simply related~by
\be
\label{dobbandlepI}
|\phi_{\beta\beta}|=|\delta_{L}|\,.
\ee
It is worth mentioning that the result is independent of the
approximation
$|c^{\nu}_{23}m^{\nu}_{D22}|=|s^{\nu}_{23}m^{\nu}_{D22}|$, as it
provides a simply relation between phases appearing in two very
different processes.\footnote{This result has also been presented
  in~\cite{King:2002qh}.}

\paragraph{II.}
In this limit $m_{\nu_2}\gtrsim m_{\nu_1} \gg m_{\nu_3}$, thus as we
can see from eq.~(\ref{bbnu0amp}), the relevant phase for
$\beta\beta_{\nu 0}$ is $2\sigma_1=\phi_2-\phi_1$.  For the case of
inverted hierarchies for the mass matrix of eq.~(\ref{majmasseff}),
with neutrino mass matrices satisfying the
conditions~(\ref{subsequdo2}), we can see from eqs.~(\ref{massinvh})
that the phase $\phi_2-\phi_1$ is supressed by a small factor ($c^{\nu
  2}_{12}-s^{\nu 2}_{12}$), given the similarity of the two masses,
$m_{\nu_2}\gtrsim m_{\nu_1}$. Thus the main contribution to this phase
is
\be
|\phi_{\beta\beta}|=|2\sigma_1|=\left|\frac{(c^{\nu 2}_{12}
-s^{\nu 2}_{12})}{f}\sin(2\gamma^{\nu\prime}_{12})\right|,
\ee
where $f$ is a further supression factor.\footnote{
  $f^2=\sin(2\gamma^{\nu\prime}_{12})^2
  \cos(2\theta^{\nu}_{12})^2+\left[
    (3+\cos(4\theta^{\nu}_{12}))+(m^{\nu\prime 2}_{11}+m^{\nu\prime
      2}_{22})c^{\nu 2}_{12} s^{\nu
      2}_{12}/|m^{\nu\prime}_{11}m^{\nu\prime}_{22}|\right]^2 $.}  In
this case, as it can be seen from
eqs.~(\ref{gammasinvhi}),~(\ref{deltalinteta}), the relevant phase for
neutrinoless double beta decay and the phase for leptogenesis,
eq.~(\ref{deltalinteta}), are related by
\be
\label{dobbandlepII}
|\phi_{\beta\beta}|=\left|\frac{(c^{\nu 2}_{12}-s^{\nu 2}_{12})}{f} 
\sin\left(2(\phi_{21}^D-\phi_{12}^D)+\delta_{\rm L}
-2\delta^{\nu}\right)\right|,
\ee
where $\delta^{\nu}$ is determined by eq.~(\ref{chidisma2}), in the
simplest case $m^{\nu\prime}_{22}=m^{\nu\prime}_{11}$, then
$2\delta^{\nu}$ and $-\delta_{\rm L}$ cancel.  If the mass matrix of
eq.~(\ref{majmasseff}), with the conditions~(\ref{subsequdo2}), is
realized in the symmetry basis, with small mixings in the leptonic
sector, as in example II of section~\ref{majmasseff}, then we also
need to introduce the contribution of $\gamma^{l\prime}_{12}$ from the
mass matrix of charged leptons.

For large values of $\tan\beta$ ($\geq 20$) order 1 phases $\phi_i$
experience a small increase ($\sim 2\%$) in their values at
electroweak scale, $M_{\mrm{EW}}$, with respect to their values at GUT
scale, $M_{\mrm{G}}$. For small values of $\phi_i$, the increase on
the phases can be as large as $\sim 50\%$. For small values of
$\tan\beta$ and for all the values of $\phi_i$, the effect of the
RGE's on these phases is small ($\lesssim 5\%$).  In the flavour
basis, there are two contributions to the change in the Dirac CP
violating phase $\delta$ with respect to $t=\ln(\mu/\mu_O)$,
$d\delta/dt$~~\cite{Antusch:2003kp}, one is proportional to
$m_{\nu_1}m_{\nu_3}\sin(\phi_1-\delta)$ thus if $m_{\nu_3}$ is
negligible then this contribution is sub-dominant. In this case, the
other contribution, proportional to
$m_{\nu_1}m_{\nu_2}\sin(\phi_1-\phi_2)$, becomes the relevant one, as
is the case of the inverted hierarchy presented in
section~\ref{InvHiers}. Hence, if $\phi_1-\phi_2$ does not change
significatively, the same happens to $\delta$. In the flavour basis,
$\delta$ is given by eq.~(\ref{chidisma2}), which is the same
combination relevant for neutrinoless double beta decay,
eq.~(\ref{dbdih2fv}), thus the effects of the RGE's on $\delta$ are
the same that $\phi_1-\phi_2$ experiences. In the symmetry basis, when
the contribution $\gamma^l_{12}$ has to be taken into account there is
not a significative change in $\delta$ because of the hierarchy in
$Y^l$ and the weak effects of the RGE's on it. The RGE's produce an
increase of $\lesssim 30\%$ in the mass scale of the neutrino masses
at $M_{\mrm{EW}}$ with respect to $M_{\mrm{G}}$.

\section{Summary and outlook}\label{section7}

In order to identify probable symmetries underlying the charged
leptons and neutrinos, it is useful to work in the symmetry basis, in
which the pattern of the possible (broken) symmetries underlying the
leptons are realized, and for which both the neutrinos and leptons
mass matrices may not be diagonal. In this way, in general, we can
study the contributions that the elements of their diagonalization
matrices give to the parameters of the $U_{\rm MNS}$ matrix.  We have
motivated two successful hierarchies for neutrinos,
eq.~(\ref{subsequdo}) and eq.~(\ref{subsequdo2}), through the family
symmetries, $\SU(3)_F$ and $\U(1)_F$ respectively, and studied the
contribution from the mixings diagonalizing the charged leptons to the
elements of $U_{\rm MNS}$, in particular to the Dirac CP violation
phase, $\delta_{\rm O}$, which will be measured in neutrino
oscillation experiments.

The contribution to the elements of $U_{\rm MNS}$ coming from the
diagonalization of charged leptons may save some of the patterns
considered to reproduce the observed mass splittings and mixing angles
for neutrinos, eq.~(\ref{expcons}) (which give for example a nearly
exact maximal mixing explaining solar neutrinos experiments), in the
sense that can receive contributions from the charged leptons. As we
have seen, the angle $\theta_{\rm sol}$ may receive contributions from
$\theta^l_{12}$ which can help $\theta_{\rm sol}$ to deviate from
maximality. The same happens with the element $\theta_{\rm rct}$ which
can be increased or decreased by taking into account contributions
from $\theta^l_{12}$.

A direct relation between the phases appearing in leptogenesis and
neutrino oscillation, in general, does not exist. However given that
leptogenesis is a very attractive mechanism to produce the baryon
asymmetry observed in the universe, which is measured to a high
precision, eq.~(\ref{ybmeas}), it is worth-while to look for a
connection in models which can describe correctly the observed
neutrino mass splittings and mixings.  In the leptogenesis scenario
the sign of the lepton asymmetry, $Y_L$, is fixed to reproduce a
positive baryon asymmetry, $Y_B$. If we can write the terms appearing
in $Y_L$, namely $\varepsilon_j$ -the asymmetry produced by the decay
of the heavy right Majorana neutrinos-, in terms of parameters
appearing in neutrino oscillations then it is interesting to determine
whether or not there is a relation among the phases appearing in these
processes and also if the relative sign of $\varepsilon_j$ and
$J_{CP}$ may be determined.

We have seen that for hierarchies, eqs.~(\ref{subsequdo}) and
eq.~(\ref{subsequdo2}), describing the mass terms for the low energy
neutrinos, $m^{\nu}_{LL}$, there are interesting relations among
phases appearing in CP violation for neutrino oscillations and
leptogenesis. In cases like these, the phase $\delta_{\rm O}$, when
measured in future neutrino oscillation experiments, will help to
constraint the possible patterns for neutrino matrices and will
determine whether or not these patterns are fully compatible with the
leptogenesis, in the sense that they could be able to reproduce both
the magnitude and the sign of the baryon asymmetry in the universe,
$Y_B$.

Although the sensitivity of experiments involving neutrinoless double
beta decay processes needs to be further increased, we can determine
the predictions for the Majorana phases from the models considered
here. We have also studied the relations between the Majorana phases
and phases appearing in leptogenesis; it is remarkable that in some
cases these relations are simple, eqs.~(\ref{dobbandlepI}).

Given a successful model describing the masses for leptons, it is
interesting to look for predictions relevant to leptogenesis and
neutrinoless double beta decay parameters. For example, one can try to
identify a possible symmetry to describe the hierarchy $IH1$,
eq.~(\ref{plausinvhi}). This is the pseudo-Dirac limit, for which
$m_{\nu_1}=-m_{\nu_2}$, and hence the relevant phase for
$\beta\beta_{\nu O}$ decay becomes trivial, $2\sigma_1=2\pi$.

\acknowledgments
I would like to thank G. G. Ross, F. A. Dolan and N. T. Leonardo for
useful discussions and careful reading of the manuscript, S. King,
J. W. Valle and O. Vives for very useful discussions; S. Parameswaran
and E. I. Zavala for encouragement and detailed reading of the
manuscript and finally to CONACyT-Mexico for financial support.

\appendix

\section{Diagonalization of hierarchical mass matrices}\label{appA}

\subsection{Diagonalization of the hermitean matrix $H=mm^{\dagger}$}
\label{Diagofhiermassm}\label{section8.1}

Writing $m$ in the form $m_{i,j}=|m_{i,j}|e^{i\phi_{ij}}$, the
hermitean matrix $H=mm^{\dagger}$ is given by:
\be
\label{HermiteMat}
H=mm^{\dagger}=\pmatrix{
H_{11}& H_{12}e^{-i \gamma_{12}} &H_{13}e^{-i \gamma_{13}}\cr
H_{12}e^{i \gamma_{12}}&H_{22}&H_{23}e^{-i \gamma_{23}}\cr
H_{13}e^{i \gamma_{13}}&H_{23}e^{i \gamma_{23}}&H_{33}},
\ee
where 
\bea
\label{Gammaij}
H_{1j}e^{i\gamma_{1j}}&=& m^*_{11}m_{j1}+m^*_{12}m_{j2}+m^*_{13}m_{j3}
\nonumber\\ 
H_{23}e^{i\gamma_{23}}&=& m^*_{21}m_{31}+m^*_{22}m_{32}+m^*_{23}m_{33}\,.
\eea
For the case of quarks and charged leptons ($f=u,d,l$) we use
$H^{f}=m^f m^{f\dagger}$, for the case of the effective mass of
neutrinos, in the notation of eq.~(\ref{seesawfor}), we use
\be
H^{\nu}=m^{\nu\dagger}m^{\nu}\,,
\ee
consequently the elements of $H^{\nu}$ are as in eq.~(\ref{Gammaij}),
with the replacements $H^{\nu}_{ij}e^{\gamma^{\nu}_{ij}}$
$\rightarrow$ $H_{ij}e^{-\gamma_{ij}}$.  $H$ can be diagonalized
through rotations and re-phasings which at the end of the procedure
can be written in terms of just three rotation angles and six phases,
since it is a hermitean matrix, but only three of these are fixed by
the elements of $H$. For hierarchical mass matrices, the first
rotation should be in the sector for which a rotation is big (i.e.\
$\tan$ of rotation of order 1), then we can pull out some of the
phases and continue making rotations until the off diagonal elements
are negligible in comparison with the diagonal ones.

Let us begin with a rotation plus a re-phasing in the 23 sector,
making the entries 23 and 32 of $H$ zero:
\bea
H^{\prime}&=&V^{\dagger}_{23}HV_{23}
\\
&=&\pmatrix{
H_{11}^{\prime}& H_{12}^{\prime}e^{-i \gamma^{\prime}_{12}} 
&H_{13}^{\prime}e^{-i \gamma^{\prime}_{13}}\cr
H_{12}^{\prime}e^{i \gamma^{\prime}_{12}}&H_{22}^{\prime}&0\cr
H_{13}^{\prime}e^{i \gamma^{\prime}_{13}}&0&H_{33}^{\prime}},
\qquad\!
V_{23}=
\pmatrix{
1&0&0\cr
0&1&0\cr
0&0&e^{i \gamma_{23}}}
\! \pmatrix{
1&0&0\cr
0&c^f_{23}&s^f_{23}\cr
0&-s^f_{23}&c^f_{23}}.
\nonumber
\eea
Now  we can extract the phases  as follows:
\be
\label{Hafter1step}
H^{\prime}=\pmatrix{
e^{-i \gamma^{\prime}_{12}}&0&0\cr
0&1&0\cr
0&0&e^{i (\gamma^{\prime}_{13}-\gamma^{\prime}_{12})}}
\!\pmatrix{
H^{\prime}_{11}& H^{\prime}_{12}&H^{\prime}_{13}\cr
H^{\prime}_{12}&H^{\prime}_{22}&0\cr
H^{\prime}_{13}&0&H^{\prime}_{33}}
\!\pmatrix{
e^{i \gamma^{\prime}_{12}}&0&0\cr
0&1&0\cr
0&0&e^{-i(\gamma^{\prime}_{13}-\gamma^{\prime}_{12})}}
=P^{\dagger}H^{\prime}_r P\,,
\ee
and we can absorb the phases appearing in $H^{\prime}$ in the
definition of the matrix $V_{23}$ by defining
$V_{23}^{\prime}=V_{23}{\rm diag}(e^{-i \gamma^{\prime}_{12}},1,e^{i
  (\gamma^{\prime}_{13}-\gamma^{\prime}_{12})})$.  We note that if
$H^{\prime}_{13}\ll H^{\prime}_{33}$ we may continue rotating the
matrix $H^{\prime}_r$ by a rotation in the 13 sector whose angle
$\theta_{13}\approx H^{\prime}_{13}/H^{\prime}_{33}$ will be small and
thus we have
\be
H^{\prime\prime}_r=R^t_{13}H^{\prime}_rR_{13}=
\pmatrix{
H^{\prime\prime}_{11}&H^{\prime\prime}_{12}&0\cr
H^{\prime\prime}_{22}&H^{\prime}_{22}&s_{13}H^{\prime}_{12}\cr
0& s_{13}H^{\prime}_{12}&H^{\prime\prime}_{33}}.
\ee
If $(s_{13}H^{\prime}_{12})$ is negligible with respect to the other
elements of the matrix then we can continue with a rotation, $R_{12}$
in the 12 sector, which will produce an approximate diagonal matrix if
also $(s_{12}H^{\prime}_{13})$ is negligible with respect to the other
elements. Thus the matrix $H$ would be diagonalized only through three
successive rotations and re-phasings, which we can identify
immediately with the required three mixing angles and phases required
to diagonalize any hermitean matrix:
\be
\label{Hdiagap}
H_{{\rm diag}}= R^{t}_{12} R^{t}_{13} V^{\prime \dagger}_{23}
H V^{\prime}_{23} R_{13} R_{12}\,,
\ee
where we define $L^{\dagger}\equiv R^{t}_{12} R^{t}_{13} V^{\prime
  \dagger}_{23}$.

\pagebreak[3] 

In this approximation the tangents of the diagonalization angles are given by
\bea
\label{tanthetaij}
t(2\theta^f_{23})&\approx&\frac{2H_{23}}{H_{33}-H_{22}}\,,
\nonumber\\
t(2\theta^f_{13})&\approx&\frac{2H^{\prime}_{13}}{H^{\prime}_{33}
-H^{\prime}_{11}}=\frac{2|s^f_{23}H_{12}e^{-i\gamma^f_{12}}
+e^{i(\gamma^f_{23}-\gamma^f_{13})}c^f_{23}H_{13}|}{c^{f2}_{23}
[H_{33}+t^{f2}_{23}H_{22}+2t_{23}H_{23}]-H_{11}}\,,
\nonumber\\
t(2\theta^f_{12})&\approx&\frac{2H^{\prime\prime}_{12}}{
H^{\prime\prime}_{22}-H^{\prime\prime}_{11}}
\\
&=&\frac{2c^f_{13}|H_{12}c^f_{23}e^{-i\gamma^f_{12}}-e^{i(\gamma^f_{23}
-\gamma^f_{13})}s^f_{23}H_{13}|}{c^f_{23}|H_{22}{+}t^{f2}_{23}
H_{33}{-}2t^f_{23}H_{23}|{-}[c^{f2}_{13}H_{11}{-}2c^f_{13}s^f_{13}|
c^f_{23}H_{13}e^{i(\gamma^f_{23}{+}\gamma^f_{12}{-}\gamma^f_{13})}{+}H_{12}
s^f_{23}|]}\,,
\nonumber
\eea
where we have also made explicit the relationship between the elements
of $H^{\prime\prime}$, $H^{\prime}$ and $H$, and hence we can identify
the phases $\gamma^{f\prime}_{13}$ and $\gamma^{f\prime}_{12}$,
\bea
\label{gammapridef}
\gamma^{f\prime}_{13}&=&{\rm Arg}[c^{f}_{23}H_{13}e^{i(\gamma^f_{13}
-\gamma^f_{23})}+s^{f}_{23}H_{12}e^{i\gamma^f_{12}}]\,,
\nonumber\\
\gamma^{f\prime}_{12}&=&{\rm Arg}[c^{f}_{23}H_{12}e^{i\gamma^f_{12}}
-s^{f}_{23}H_{13}e^{i (\gamma^f_{13}-\gamma^f_{23})}]\,.
\eea

The procedure presented above may be used when diagonalizing canonical
or inverted hierarchies of neutrinos (and some hierarchies of charged
leptons and quarks), satisfying the following conditions for (strong)
\emph{Canonical hierarchies}, $m^f_1\ll m^f_2\ll m^f_3$,
\be
\label{condcanhierD}
m^f_{33}={\mathcal{O}}(m^f_{22},m^f_{23})\,,
\quad 
m^f_{33}\gg{\mathcal{O}}(m^f_{13})\,,
\quad
m^f_{13}={\mathcal{O}}(m^f_{12})\,,
\quad 
m^f_{11}\ll {\mathcal{O}}(m^f_{12}, m^f_{23}, m^f_{33})\,,
\ee
or for the case of \emph{Inverted hierarchies}, $m^f_1,m^f_2\gg
m^f_3$, such that
\be
\label{condinvhierD}
m^f_{33}={\mathcal{O}}(m^f_{22},m^f_{23},m^f_{11}),
\qquad 
m^f_{33}\ll {\mathcal{O}}m^f_{13}\,,
\qquad
m^f_{13}={\mathcal{O}}(m^f_{12})\,.
\ee
We remark that for both cases the hermitean matrix,
$H^f=m^fm^{f\dagger}$, has the structure:
$H^f_{22}={\mathcal{O}}(H^f_{33})$,
$H^f_{33}\gg{\mathcal{O}}(H^f_{13})$,
$H^f_{13}={\mathcal{O}}(H^f_{12})$. However due to the different
hierarchies the tangents of the mixing angles will be dominated by
different elements of the original matrices, $m$, as can be seen in
appendix~\ref{appB}.

\subsection{ Diagonalization of $H$ vs. diagonalization of $m$}\label{section8.2}

If $H$ is diagonalized by $L$: $H_{\rm diag}=L^{\dagger}H L$,
eq.~(\ref{Hdiagap}), then $L$ can be written as
\be
\label{LH}
L=\pmatrix{
1&0&0\cr
0&1&0\cr
0&0&e^{i \gamma_{23}}}
R_{23}
\pmatrix{
1&0&0\cr
0&e^{i \gamma^{f\prime}_{12}}&0\cr
0&0&e^{i \gamma^{f\prime}_{13}}}
R _{13}R _{12}
e^{-i\gamma^{f\prime}_{12}}\,,
\ee
on the other hand if $m$ is diagonalized by $L_m$, i.e.\ $m_{\rm
  diag}=L^{ t}_m mL^{}_m$, can be expressed in general by
\bea
L_m&=&\pmatrix{
1&0&0\cr
0&e^{i\beta_2}&0\cr
0&0&e^{i\beta_1}
}
R _{23}
\pmatrix{
1&0&0\cr
0&e^{i\beta_3}&0\cr
0&0&1}
R _{13}R _{12}\pmatrix{
e^{-i\alpha_0}&0&0\cr
0&e^{-i\alpha_1}&0\cr
0&0&e^{-i\alpha_2}}
\nonumber\\
&=&P_1R _{23}P_2R _{13}R _{12}P_3\,.
\eea
Re-arranging the phases we have
\be
\label{Lh}
L_m=\pmatrix{
1&0&0\cr
0&1&0\cr
0&0&e^{i(\beta_1-\beta_2)}}
R _{23}\pmatrix{
1&0&0\cr
0&e^{i(\beta_2+\beta_3)}&0\cr
0&0&e^{i\beta_2}}
R _{13}R _{12}P_3\,.
\ee
Comparing eq.~(\ref{Lh}) with eq.~(\ref{LH}) it can be seen that $L_m
P^{\dagger}_3=L$, thus we have the following relations:
\be
\gamma _{23}=\beta_1-\beta_2\,, 
\qquad 
\gamma^{\prime}_{12}=\beta_2+\beta_3\,, 
\qquad 
\gamma^{\prime}_{13}=\beta_2\,.
\ee
Hence we can express the matrix $L_m$ diagonalizing the mass matrix
$m$ in terms of elements of the diagonalization of $H$ and so express
the $U_{\rm MNS}$ matrix in these terms. It is useful to notice that
$L_m$ can be rewritten as follows
\bea
\label{diagmatLm}
L_m&=&P(1,e^{i\gamma^{\prime}_{12}},e^{i(\gamma^{\prime}_{12}
+\gamma _{23})})R _{23}P(e^{i(\gamma^{\prime}_{12}
-\gamma^{\prime}_{13})},1,1)R_{13}P(e^{-i(\gamma^{\prime}_{12}
-\gamma^{\prime}_{13})},1,1)R _{12}\times
\nonumber\\ 
&&\times P(1,1, e^{i(\gamma^{\prime}_{12}-\gamma^{\prime}_{13})})P_3\,,
\eea
because if we were working in the flavour basis then $U_{\rm
  MNS}=L^{\nu}_m$, and hence we can identify the CP violation phase
appearing in neutrino oscillations by comparing to the standard
parameterization~(\ref{stparmixm}) so that
\be
\delta_{\rm O}=\gamma^{\nu\prime}_{13}-\gamma^{\nu\prime}_{12}\,.
\ee
We can write the diagonal $m_{\rm diag}$ matrix with phases as:
\be
m_{\rm diag}=L^t_m m L_m 
={\rm diag}\left(e^{-2i\alpha_1}m_1,e^{-2i\alpha_2} m_2,  
e^{-2i\alpha_3} m_3 \right),
\ee
where the $\alpha_i$'s make the eigenvalues $m_{\rm diag}$ real, i.e.\
$\alpha_i=\phi_i/2$ and in the case of neutrinos the two Majorana
phases $\sigma_1$ and $\sigma_2$ can be identified as follows
\be
\sigma_1=2(\alpha_2-\alpha_1)=(\phi_1-\phi_2)\,,
\qquad 
\sigma_2-\delta_2=2(\alpha_2-\alpha_3)=(\phi_3-\phi_2)
\ee
in the convention $P(\sigma)={\rm
  diag}(e^{i\sigma_1},1,e^{i\sigma_2})$, as in eq.~(\ref{convmajph}).

\section{Masses, mixing angles and  phases}\label{appB}

\subsection{Canonical hierarchies}\label{section9.1}

\subsubsection{Masses}\label{section9.1.1}

For both cases, canonical and inverted hierarchies, the diagonal
masses $m^{\nu}_i=|m^{\nu}_i|e^{i\phi_i}$ are obtained by $L_m^t m
L_m$, eq.~(\ref{Lh}), for the case of canonical hierarchies under the
conditions~(\ref{subsequdo}) we have the following approximations, in
terms of the elements of the Dirac mass matrix and the heavy Majorana
mass matrix,
\be
m_{\nu_2}\approx  \frac{|m^{\nu}_{D12}|^2}{M_2 \leb s^{\nu}_{12}\rib^2}\,, 
\qquad
m_{\nu_3}\approx  \frac{|m^{\nu}_{D21}|^2+|m^{\nu}_{D31}|^2}{M_1}\,.
\ee
In this approximation, the phases of these elements are given by
\be
\label{phasesi33}
\phi_2\approx 2\phi^D_{12}\,,
\qquad
\phi_3\approx -2\gamma^{\nu\prime}_{13}+2\phi^D_{21}\,.
\ee

\subsubsection{Tangents of the mixing angles}\label{section9.1.2}

The angle of the rotation $R_{23}$ in terms of the elements of
$H^{\nu}$ is given by
\be
\label{t2theta23H}
t(2\theta^f_{23})=\frac{2H_{23}}{H_{33}-H_{22}}=\frac{|m_{13}m^{*}_{12}
+m_{23}m_{22}^{*}+m_{33}m_{32}^*|}{|m_{33}|^2-|m_{22}^2+|m_{13}|^2
-|m_{12}|^2+|m_{23}|^2-|m_{32}|^2}
\ee 
where the last term in the denominator vanishes for symmetric matrices.

For the case of canonical hierarchies such that $|m_{12}||m_{11}|$
$\ll |m_{2i}||km_{1i}|$ and $|m_{22}||m_{12}|$
$={\mathcal{O}}(|m_{23}||m_{13}|)$, the same expression as
eq.~(\ref{t2theta23H}) is obtained for $t(2\theta^f_{23})$ if we begin
diagonalizing the matrix $m$ by performing the re-phasing with the
diagonal matrix of phases $P_2$ and the $R^f_{23}$ rotation:
\be
m^{\prime}=R^f_{23}P_2^t m P_2R^f_{23}\,.
\ee
This happens because in this case the condition that make zero the
entries (23) and (32)~is:
\be
t(2\theta^f_{23})=\frac{2|m_{23}|}{|m_{33}|e^{i(\phi_{33}-\phi_{23}
+(\beta_1-\beta_2))} -|m_{22}|e^{i(\phi_{22}-\phi_{23}-(\beta_1-\beta2))} }
\ee
and requiring this quantity to be real, i.e.\ that 
\be
|m_{33}|\sin(\phi_{33}-\phi_{23}+(\beta_1-\beta2))=|m_{22}|
\sin(\phi_{22}-\phi_{23}-(\beta_1-\beta2))\,,
\ee
is equivalent to
\be
\label{equivg23betas}
\tan(\beta_2-\beta_1)=\frac{|m_{22}|\sin(\phi_{23}-\phi_{22})
+|m_{33}|\sin(\phi_{33}-\phi_{23})}{|m_{22}|\cos(\phi_{23}
-\phi_{22})+|m_{33}|\cos(\phi_{33}-\phi_{23})}
\ee
which is the same as equation~(\ref{Gammaij}) for
$\gamma_{23}=(\beta_1-\beta_2)$ and the kind of hierarchies of
eq.~(\ref{subsequdo}).

The same kind of equivalences, among the rest of the elements of the
diagonalization of $H$ (angles and phases involved in it) and the
elements of the diagonalization of $m$, apply, considering the
hierarchies of eq.~(\ref{subsequdo}).  The advantage of diagonalizing
$H=m m^{\dagger}$ is that we can extract the relevant phases for CP
violation after the first step of diagonalization
eq.~(\ref{Hafter1step}).

We present here approximate formulas for the angles diagonalizing the
effective neutrino mass matrix in terms of the elements of the
neutrino Dirac matrix for the case of hierarchies presented in
section~\ref{predcanhier}.  At leading order
\bea
\label{tangentsang}
t^{\nu}_{23}&\approx&\frac{|m^{\nu}_{D21}|}{|m^{\nu}_{D31}|}\,, 
\qquad 
t^{\nu}_{13}\approx\frac{M_1}{M_2}s^{\nu}_{23}\frac{|m^{\nu*}_{D12}
m^{\nu}_{D22}+m^{\nu}_{D13}m^{\nu*}_{D23}|}{|m^{\nu}_{D12}|^2
+|m^{\nu}_{D13}|^2}\,,
\nonumber\\ 
t^{\nu}_{12}&\approx&\frac{|m^{\nu}_{D12}|}{c^{\nu}_{23}|
m^{\nu}_{D22}|\cos(\phi_{22}-\phi_{12}-\gamma^{\nu\prime}_{12})
-s_{23}^{\nu}|m^{\nu}_{D23}|\cos(\phi_{32}-\phi_{12}-
\gamma^{\nu}_{23}-\gamma^{\nu\prime}_{12})}\,.\quad
\eea

\subsection{Inverted hierarchies}\label{section9.2}

The inverted Hierarchies for the effective neutrino mass matrix
$m^{\nu}_{LL}$, of the form of eq.~(\ref{plausinvhi}), produce the
following hermitean matrices
($H^{\nu}=m^{\nu\dagger}_{LL}m^{\nu}_{LL}$), respectively:
\be
{\mrm IH1}:\quad
H{=}\pmatrix{
1+\varepsilon^2&\frac{3\varepsilon}{\sqrt{2}}&\frac{3\varepsilon}{\sqrt{2}}\cr
\frac{3\varepsilon}{\sqrt{2}}&\frac{1}{2}+2\varepsilon^2 &\frac{1}{2}
+2\varepsilon^2\cr
\frac{3\varepsilon}{\sqrt{2}}&\frac{1}{2}+2\varepsilon^2 &\frac{1}{2}
+2\varepsilon^2},
\quad\,\,
{\mrm IH1}:\quad
H{=}\pmatrix{
1+2\varepsilon^2 & 2\varepsilon &2\varepsilon \cr
2\varepsilon & \frac{1}{2}+\varepsilon^2 &\frac{1}{2}+\varepsilon^2\cr
2\varepsilon&\frac{1}{2}+\varepsilon^2 &\frac{1}{2}+\varepsilon^2}.
\ee
These matrices satisfy the conditions for the diagonalization process
as outlined in section~\ref{Diagofhiermassm}. We can diagonalize them
with a diagonalization matrix $L_m$ of the form eq.~(\ref{diagmatLm}),
except that now
\be
\beta_1=\gamma^{\nu}_{23}+\gamma^{\nu\prime}_{13}+\pi\,,
\qquad 
\beta_2=\gamma^{\nu\prime}_{13}\,, 
\qquad 
\beta_3=\gamma^{\nu\prime}_{12}-\gamma^{\nu\prime}_{13}\,.
\ee
Let us take the case the case of the hierarchy $HI2$, assigning phases
of the form $\phi^{\nu}_{ij}$ and writing $m^{\nu}=m^{\nu}_{LL}$. In
this case ${\mathcal{O}}(m^{\nu}_{33})=m^{\nu}_{22},m^{\nu}_{23}$,
${\mathcal{O}}(m^{\nu}_{33})\ll m^{\nu}_{13}$ and
${\mathcal{O}}(m^{\nu}_{13})\ll m^{\nu}_{12}$. The dominant terms in
determining $\tan\theta^{\nu}_{23}$ are $|m^{\nu}_{13}m^{\nu
  *}_{12}|$, so that
\be
\tan(2\theta^{\nu}_{23})=\frac{|m^{\nu}_{13}
m^{\nu *}_{12}|}{|m^{\nu }_{12}|^2-|m^{\nu }_{13}|^2}\,, 
\ee
whose solution for $\tan\theta^{\nu}_{23}$ is
\be
\tan\theta^{\nu}_{23}=\frac{|m^{\nu}_{13}|}{|m^{\nu}_{12}|}\,.
\ee
In this case we can also obtain the phases $\gamma^{\nu}_{ij}$ from
equation~(\ref{Gammaij}) and the hierarchical conditions of $HI2$,
eq.~(\ref{plausinvhi}). The easiest phase to obtain is
$\gamma^{\nu}_{23}$ because it can be determined in the first step of
the diagonalization:
\be
\gamma^{\nu}_{23}\approx \phi^{\nu}_{13}-\phi^{\nu}_{12}\,.
\ee
After this we can continue with the diagonalization in the sectors
$13$ and $12$.  In section~\ref{InvHiers} we have presented the
results of this diagonalization in terms of the elements of the Dirac
neutrino mass matrix and the right-handed neutrino mass matrix.

\end{document}